\documentclass[aps,prl,amsmath,amssymb,superscriptaddress,twocolumn,color,epsfig,graphicx,bm,floatfix]{revtex4-1}

\usepackage{color}
\usepackage{graphicx}
\usepackage{graphics}
\usepackage{subfigure}
\usepackage{epsfig}
\usepackage{amsmath}
\usepackage{array}
\usepackage{amssymb}
\usepackage{graphicx}      
\usepackage{dcolumn}       
\usepackage{bm}            
\usepackage{units}
\usepackage{array}
   
\begin{document}

\title{Low-density silicon allotropes for photovoltaic applications}

\author{Maximilian Amsler}
\affiliation{Department of Physics, Universit\"{a}t Basel,
Klingelbergstr. 82, 4056 Basel, Switzerland}
\author{Silvana Botti}
\affiliation{Institut f\"{u}r Festk\"{o}rpertheorie und -optik, Friedrich-Schiller-Universit\"{a}t Jena and European Theoretical Spectroscopy Facility,
Max-Wien-Platz 1, 07743 Jena, Germany}
\author{Miguel A.L. Marques}
\affiliation{Institut f\"{u}r Physik, Martin-Luther-Universit\"{a}t Halle-Wittenberg, D-06099 Halle, Germany}
\author{Thomas J. Lenosky}
\affiliation{1974 Kirby Way, San Jose California 95124-1324, USA}
\author{Stefan Goedecker}
\email{Stefan.Goedecker@unibas.ch}
\affiliation{Department of Physics, Universit\"{a}t Basel,
Klingelbergstr. 82, 4056 Basel, Switzerland}

\date{\today}

\begin{abstract}

Silicon materials play a key role in many technologically relevant fields, ranging from the electronic to the photovoltaic industry. A systematic search for silicon allotropes was performed by employing a modified \textit{ab initio} minima hopping crystal structure prediction method. The algorithm was optimized to specifically  investigate the hitherto barely explored low-density regime of the silicon phase diagram by imitating the guest-host concept of clathrate compounds. In total 44 metastable phases are presented, of which 11 exhibit direct or quasi-direct band-gaps in the range of $\approx$~1.0-1.8~eV, close to the optimal Shockley-Queisser limit of $\approx 1.4$~eV, with a stronger overlap of the absorption spectra with the solar spectrum compared to conventional diamond silicon. Due to the structural resemblance to known clathrate compounds it is expected that the predicted phases can be synthesized.
\end{abstract}

\pacs{somepacs}

\maketitle

\section{Introduction}

Silicon has become one of the technologically most relevant materials in the last century. It is the second most abundant element on earth and thus readily available, rendering it ideal for many industrial applications. Since elemental silicon is an intrinsic semiconductor that can be both $p$ and $n$-doped it is well suited and thus widely used in the electronic industry, with applications in integrated circuits, photovoltaics and optoelectronic devices. Due to its dominance in electronics, it has been extensively studied both experimentally and theoretically such that its properties in the most stable form, the diamond structure ($d$-Si), are well understood.

Although diamond silicon dominates the photovoltaic market due to its low cost, it is in fact not a good photo absorber. The band-gap of 1.12~eV in $d$-Si~\cite{bludau_temperature_1974} is close to the maximal efficiency in the Shockley-Queisser limit of $\approx 1.4$~eV~\cite{shockley_detailed_1961}. However, it is indirect with a large difference to the direct gap of 3.2~eV such that electron excitations in the optical regime need to be phonon-mediated~\cite{lautenschlager_temperature_1987}, preventing the use of $d$-Si in thin-film solar cells or other optical applications such as light-emitting diodes or high-performance transistors. Therefore, efforts have been made to design, discover and synthesize direct band-gap materials with improved absorption properties.

The phase diagram of silicon is complex with a sequence of structures appearing in the high pressure regime~\cite{duclos_experimental_1990,mcmahon_new_1993} such that several metastable phases of silicon have been considered for use in photovoltaic applications~\cite{malone_prediction_2012,malone_electronic_2010,malone_emph_2008,malone_emph_2008-1}. Allotropes such as the BC8, R8 or londsdaleite silicon however turn out to be unsuited since they are either semi-metallic (BC8), have a too small (0.24~eV for R8) or indirect band-gaps ($\approx$1~eV for londsdaleite)~\cite{besson_electrical_1987,malone_emph_2008}. Recently, novel silicon allotropes have been proposed and studied theoretically through structural search methods, such as the Si20T~\cite{xiang_towards_2013} and P-1~\cite{botti_low-energy_2012} phases, which exhibit promising electronic properties with either direct or quasi-direct band-gaps in the desired energy range. Carefully tailored super-lattices of interlaced silicon and germanium layers were also proposed by D'Avezac~\textit{et al.}~\cite{davezac_genetic-algorithm_2012} with excellent absorption, but techniques for the required atomic scale layer-by-layer assembly on an industrial scale are yet to be developed.

On the other hand, the low-density regime of the silicon phase diagram remains relatively poorly explored, although semi-conducting group IV clathrates~\cite{rowe_thermoelectrics_2005,beekman_inorganic_2008,kovnir_semiconducting_2004} have recently drawn attention as a possible material for various energy applications. Since the structure of clathrates is characterized by face-sharing cage-like units of host atoms enclosing guests, its properties can be tuned through tailored substitution of the host or guest elements. Superconductivity was observed in clathrate compounds~\cite{kawaji_superconductivity_1995,li_superconductivity_2013} with transition temperatures as high as 8~K in the barium-encapsulated type I silicon clathrate Ba$_8$Si$_{46}$ synthesized at high pressure~\cite{yamanaka_high-pressure_2000}. Furthermore, the properties induced by the vibrational ``rattling'' modes of the guest atoms in the large host cages leads to reduced thermal conductivity at high electronic conductivity (the phonon glass-electron crystal concept introduced by Slack~\cite{nolas_semiconducting_1998}), such that clathrate have been studied in the context of thermoelectric applications~\cite{nolas_thermoelectric_2001,rogl_formation_2005,san-miguel_high-pressure_2005,saramat_large_2006,connetable_structural_2007,christensen_thermoelectric_2010}. More recently, these materials have also been investigated for their potential use in lithium ion batteries~\cite{yang_silicon_2013,langer_electrochemical_2012,wagner_electrochemical_2014} and as hydrogen storage materials~\cite{okamoto_characterization_2007} due to the large voids that could be charged/discharged with guest atoms or molecules.

Guest free group IV element clathrates of type I and II have been touted as promising candidates for photovoltaic applications due to the direct band-gap character with gap energies close to the optimal value~\cite{baranowski_synthesis_2014,martinez_synthesis_2013,krishna_efficient_2014}. The type II silicon clathrate has a wide direct band-gap of 1.9~eV, which is however not allowed for electronic dipole transitions~\cite{connetable_structural_2007}. Nevertheless, Baranowski~\textit{et al.} investigated the possibility of tuning the electronic properties of the type II clathrate through variation of the chemical composition in Si-Ge alloys, resulting in band-gaps between 0.8-1.8~eV~\cite{baranowski_synthesis_2014}. Furthermore, novel experimental techniques were developed to efficiently prepare large samples of phase-pure type II silicon clathrates, an essential requirement for future large-scale industrial developments~\cite{krishna_efficient_2014}.

Although only group IV clathrates of type I, II, III, VIII and IX were observed experimentally, theoretical studies indicate that there is a large structural diversity of low-density silicon phases~\cite{zwijnenburg_extensive_2010}. Structural principles of various polytypes and intergrowth frameworks of clathrates were systematically investigated from basic and extended polyhedral building blocks by Karttunen~\textit{et al.}~\cite{karttunen_structural_2011}, while solutions to the Kelvin problem, the task of partitioning three-dimensional space into cells of equal volume with minimal area, were used to predict novel low-density silicon phases~\cite{zhao_sp3-bonded_2013}. More recently, global search methods were used to predict silicon allotropes with possible applications in optoelectronics. Wang~\textit{et al.}~\cite{wang_direct_2014} proposed 6 different metastable silicon allotropes exhibiting channel-like structures with direct or quasi-direct band gaps predicted with particle swarm optimization, and a novel cage-like structure was discovered by Nguyen~\textit{et al.}~\cite{nguyen_sp3-hybridized_2014} by a genetic algorithm. 

Experimentally, all routes to synthesize guest-free low density silicon phases involve, in some way or another, precursor materials containing silicon and guest elements. Thermal decomposition of Zintl monosilicide phases is one possible route to obtain silicon clathrates~\cite{gryko_low-density_2000,beekman_synthesis_2009}. In this process, the cations sublimate, causing a charge imbalance, and leading to the four-fold rearrangement of the silicon atoms around the remaining template cations. Other methods include chemical oxidation, spark plasma treatment of Na$_4$Si$_4$~\cite{beekman_preparation_2009} or directly from elemental Si and Na at high pressure~\cite{kurakevych_na-si_2013,kim_synthesis_2014}. The guest elements are then thermally removed, a process referred to as ``thermal degassing''. In fact, with this procedure it is possible to synthesize metastable silicon polymorphs with energies considerably beyond the ground state~\cite{kim_synthesis_2014}.

In this work we conduct a computational search for novel low-density silicon allotropes with potential use in photovoltaic applications. Using first principle calculations, we discover a plethora of hitherto unknown low-density low-energy silicon phases consisting of interlinked cage-like polyhedra and channel-like structures.

\section{Method}          

Inspired by the existence of group IV clathrates, the minima hopping crystal structure prediction method (MHM)~\cite{goedecker_2004,amsler_2010} was adapted to investigate low-density silicon allotropes. The MHM was initially designed to thoroughly scan the low-lying energy landscape of any material and to identify stable and metastable phases by performing consecutive short molecular dynamics escape steps followed by local geometry relaxations taking into account both atomic and cell variables. The potential energy surface is mapped out efficiently by aligning the initial molecular dynamics velocities approximately along soft mode directions~\cite{sicher_efficient_2011}, thus exploiting the Bell-Evans-Polanyi~\cite{roy_2009,jensen_introduction_2011} principle to steer the search towards low energy structures. The predictive power of this approach has been demonstrated in a wide range of applications~\cite{hellmann_2007,roy_2009,bao_2009,willand_2010,amsler_crystal_2012}.

The structural search was performed using the Lenosky tight-binding scheme to model the Si-Si interaction~\cite{lenosky_highly_1997}, which has shown to give sufficiently accurate results for solids at moderate computational cost~\cite{ghasemi_energy_2010,ghasemi_energetic_2014}. To imitate encapsulated guest elements, fictitious Lennard-Jones (LJ) spheres were placed in the simulation cell as inert template guest atoms. The LJ potential was shifted and truncated such that it decreases continuously to zero at a cutoff radius $r_c$ and no discontinuities arise in the forces~\cite{stoddard_numerical_1973}. To avoid the formation of LJ dimers the interaction between the LJ particles was further truncated at the minimum and shifted towards zero, resulting in a purely repulsive potential. The parameters for the LJ-silicon interaction were carefully tuned to already known silicon clathrate structures. In analogy to experimentally used inert template elements such as Na, the interaction between the LJ-sphere and the silicon framework was designed to be minimal, with a strong repulsive potential and a weak attraction. The parameters were thus set to $\sigma = 3.00$~\AA, $\epsilon=0.02$~eV, $r_c=7 \sigma $.

Density functional theory calculations were carried out to refine the structures with the projector augmented wave formalism as implemented in the VASP package~\cite{kresse_efficient_1996} using the Perdew-Burke-Erzernhof (PBE)~\cite{PBE96} exchange-correlation functional. A plane-wave cutoff energy of 500~eV was used together with a sufficiently dense k-point mesh, resulting in total energies converged to less than 1~meV/atom. Both atomic and cell variables were relaxed simultaneously until the forces on the atoms were less than 3~meV/\AA~and the stresses were less than 0.1~eV/\AA$^3$.

Electronic properties were investigated by calculating the Kohn-Sham electronic band structure within the PBE approximation to the exchange-correlation potential. Since conventional DFT systematically underestimates the band-gap, the hybrid HSE06 functional~\cite{heyd_hybrid_2003,paier_erratum:_2006,heyd_erratum:_2006} was used on a set of selected structures including $d$-Si and the type II clathrate. The band-gap error within PBE was found to lie consistently between 0.4 and 0.65~eV. Therefore, a scissor operator was subsequently used to obtain a corrected gap by shifting the unoccupied bands by an averaged value of 0.55~eV~\cite{levine_linear_1989}.

Absorption spectra were obtained within time-dependent density-functional theory~\cite{doi:10.1146/annurev.physchem.55.091602.094449,0034-4885-70-3-R02} using the code {\sc yambo}~\cite{Marini20091392} starting from PBE ground-states obtained with {\sc abinit}~\cite{Gonze20092582}. We used the semi-empirical $\alpha/q^2$ long-range exchange-correlation kernel~\cite{PhysRevLett.88.066404}, that was shown to give very good results for systems with delocalized Wannier-Mott excitons (such as the ones we expect in Si-based materials)~\cite{PhysRevLett.88.066404,PhysRevB.69.155112}. We used the (constant) value of $\alpha=-0.2$, which is the one determined for bulk silicon. For the optical spectra we used randomly shifted $k$-point grids with the number of $k$-points in direction $\alpha$ equal to $71\times|{\bf b}_\alpha|$, where ${\bf b}$ is the reciprocal lattice vector in direction $\alpha$. The number of bands included in the calculation was equal to 5 times the number of atoms, i.e., we used 3 unoccupied bands per Si atom. This amounts to a $12\times12\times12$ $k$-point mesh and 6 unoccupied bands for $d$-Si.

The phonon calculations were carried out with the frozen phonon approach as implemented in the PHONOPY package~\cite{phonopy}. Super-cells of dimension $4\times4\times4$ were used for cells with less than 12 atoms,  $3\times3\times3$  for cells with between 13 and 24 atoms, and $2\times2\times2$ super-cells for larger unit cells, which were found to give well converged results. The X-ray diffraction pattern were simulated with MERCURY~\cite{macrae_mercury_2006} with the Cu K$\alpha$ radiation of wavelength 1.54056~\AA.

\section{Results and discussion}

The MHM simulations at the tight-binding level were performed for different stoichiometries Si$_x$LJ$_y$, where $x$ and $y$ were chosen from two different approaches. First, a systematic set of stoichiometries for moderate system sizes were explored with all combinations of $x$ and $y$ in the range of $x=16,18,\dots,34$ and $y=2,4$, respectively. The initial structures were generated randomly, without any symmetry constraints. Second, a set of known structures from literature was used to generate initial structures for the MHM simulations. The energetically lowest energy clathrates of type I and II and structures of the Kelvin-type from Ref.~\cite{zhao_sp3-bonded_2013} were used. For each structure, the center of the polyhedra was located and LJ spheres were placed accordingly. The stoichiometries $[x|y]$ then resulted in $[46|8]$ from Type I, $[34|6]$ from Type II, $[12|2]$ from Si$_\textrm{K}^\textrm{I}$, $[46|8]$ from Si$_\textrm{K}^\textrm{II}$, $[82|14]$ from Si$_\textrm{K}^\textrm{III}$, $[41|7]$ from Si$_\textrm{K}^\textrm{IV}$, $[40|7]$ from Si$_\textrm{K}^\textrm{V}$, $[178|30]$ from Si$_\textrm{K}^\textrm{VI}$, $[68|12]$ from Si$_\textrm{K}^\textrm{VII}$, $[58|10]$ from Si$_\textrm{K}^\textrm{VIII}$, $[46|8]$ from Si$_\textrm{K}^\textrm{IX}$ and $[232|40]$ from Si$_\textrm{K}^\textrm{X}$.

Each MHM run resulted in several thousand structures, of which all with energies (calculated in tight-binding) 400~meV/atom higher than $d$-Si were discarded. Subsequently, the LJ spheres were removed and the remaining structures were carefully filtered for duplicates and analyzed with respect to symmetry. Due to the large size of the system under investigation, the majority of the phases encountered during the structural search were amorphous or some defect structures of a crystalline phase. In fact, less than 5\% of the phases encountered during a search had at least one symmetry element. Therefore, all structures that had no symmetry at all were discarded, unless they were very low in energy. Finally, the energetically most promising crystalline phases were refined at the DFT level. Both the atomic and cell variables were relaxed to the nearest local minimum without constraining the initial symmetry to avoid relaxation to a first or higher-order saddle point.

After the initial filtering, roughly 150 structures were considered for further analysis. Among those, several structures were rediscovered that had already been discussed in literature, such as the bonded chiral framework structure (CFS), first reported by Pickard~\textit{et al.}, which is energetically degenerate to the type II clathrate. Although the structure does not contain any cages, the volume per atom is considerably higher than $d$-Si with a value of 22.17~\AA$^3$ due to the channel-like voids. Similarly, a structure with $P4_2/mnm$ symmetry was found that was reported earlier by Nguyen~\textit{et al.}~\cite{nguyen_sp3-hybridized_2014} with an energy of 114~eV/atom above $d$-Si and a volume of 23.31~\AA$^3$. The H-clathrate, initially described by Udachin~\textit{et al.}~\cite{udachin_structure_1997}, was also found with an energy of 83~eV/atom above $d$-Si and a volume of 24.09~\AA$^3$. 

\begin{figure}[h!]
\caption{Hybrid channels embedded in diamond structure with atoms denoted by yellow (light gray) spheres. (a) Structural motif representing dimers (red or dark gray spheres) within a 8-ring channels (green or medium gray) and (b) 4-rings (red or dark gray spheres) within 10-ring channels (green or medium gray).}
\setlength{\unitlength}{1cm}
\subfigure[]{\includegraphics[height=0.5\columnwidth,angle=0]{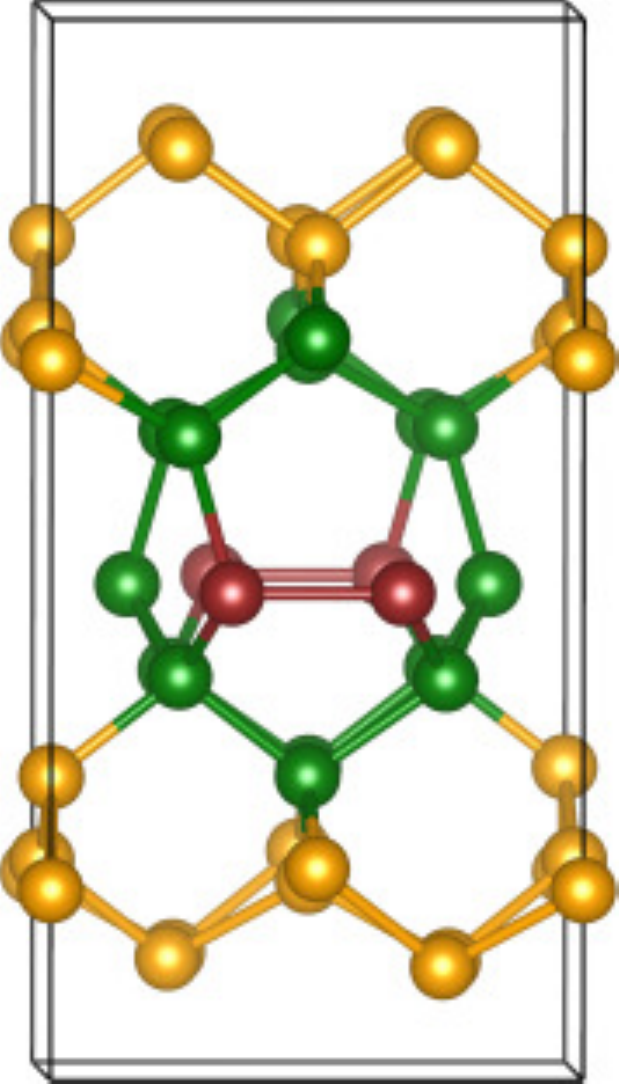}} 
\subfigure[]{\includegraphics[height=0.5\columnwidth,angle=0]{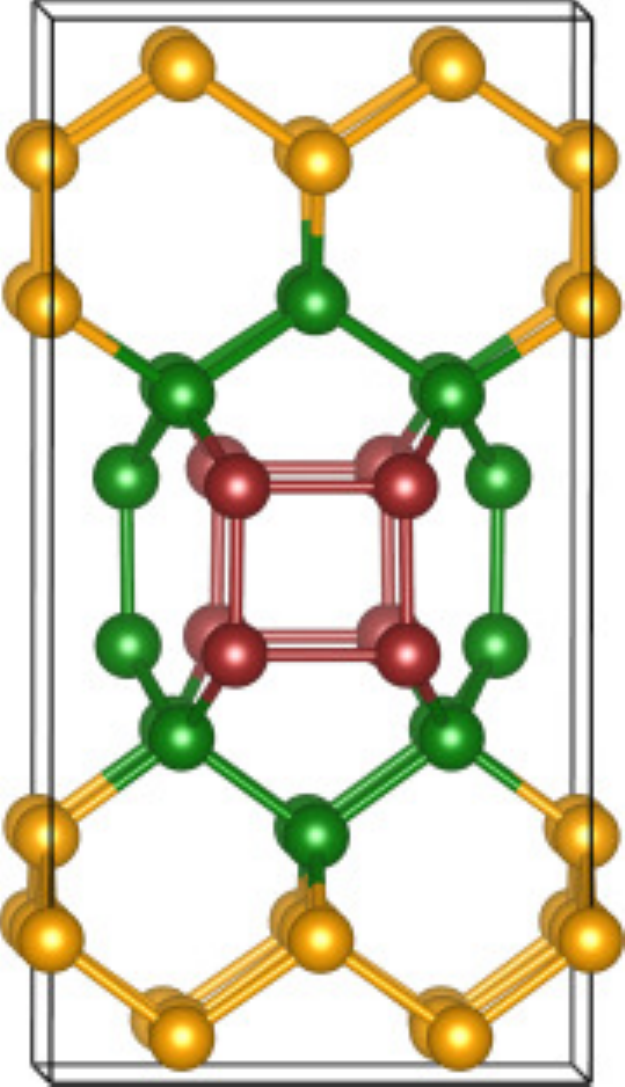}} 
\label{fig:rings}
\end{figure}

As expected, all structures are $sp^3$-bonded and have lower densities than $d$-Si due to the fictitious LJ spheres. In general, the low density in all allotropes is achieved by one of two different structural motifs. The first type consist of wide 1-dimensional channels, which either run in one, two or three dimensions, can be straight, or even run in zig-zag lines through the material. Some representatives of this class have been extensively studied in the past for various $sp^3$ materials. It has been shown that low-energy low-density allotropes with 5- to 8-membered rings can be obtained systematically from different stacking of cubic diamond layers, and structures including combinations of 4-, 6- and 8-rings have been reported in various publications~\cite{zwijnenburg_extensive_2010,amsler_crystal_2012,botti_low-energy_2012,pizzagalli_family_2013,botti_carbon_2013,bautista-hernandez_structural_2013}. Energetically, these structures are close to $d$-Si, and the channels can exist as defect wires or planes embedded in the cubic or hexagonal diamond structure~\cite{botti_carbon_2013}. The recently synthesized open framework phase of Kim~\textit{et al.}~\cite{kim_synthesis_2014}, which contains both 5- and 8-membered rings, can be classified to belong to this type and was also discovered during our search. Similar to clathrate structures, the channels are large enough to host guest atoms~\cite{kurakevych_na-si_2013}. Since these phases have been already widely discussed in literature, all structures were put into the supplementary materials. We however also discovered two other channel-like motifs which frequently appeared in low-energy allotropes during our search, namely the hybrid channels enclosing dimers within 8-rings or enclosing 4-rings within 10-rings, as shown in Figure~\ref{fig:rings}. The latter can be interpreted as chains of $[4^{2}5^{8}6^{4}_{\mathrm{III}}]$ cages within a diamond framework (see below).

\begin{figure*}[h!]
\caption{Polyhedral cages which constitute the building blocks for the crystal lattice of the novel silicon allotropes.}            
\centering
\setlength{\unitlength}{1cm}
{\includegraphics[height=0.10\textwidth]{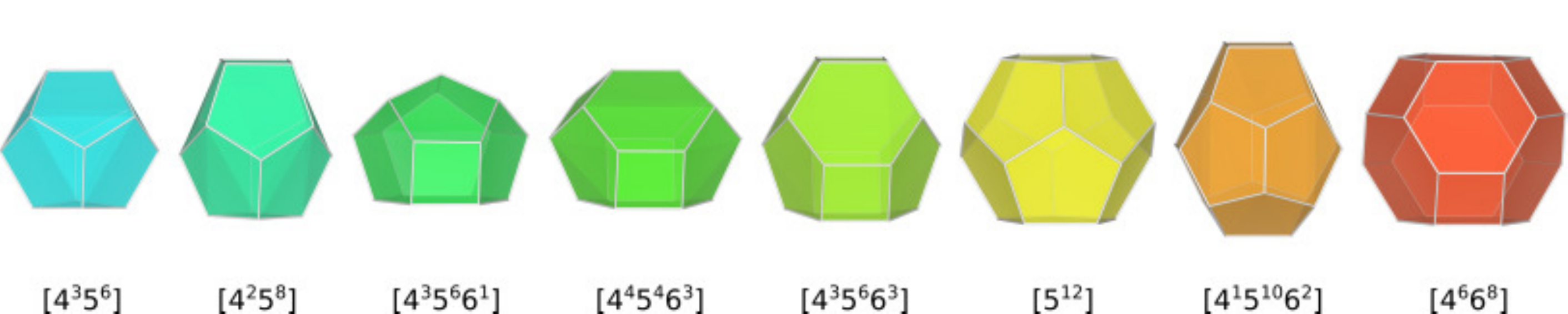}}
{\includegraphics[height=0.10\textwidth]{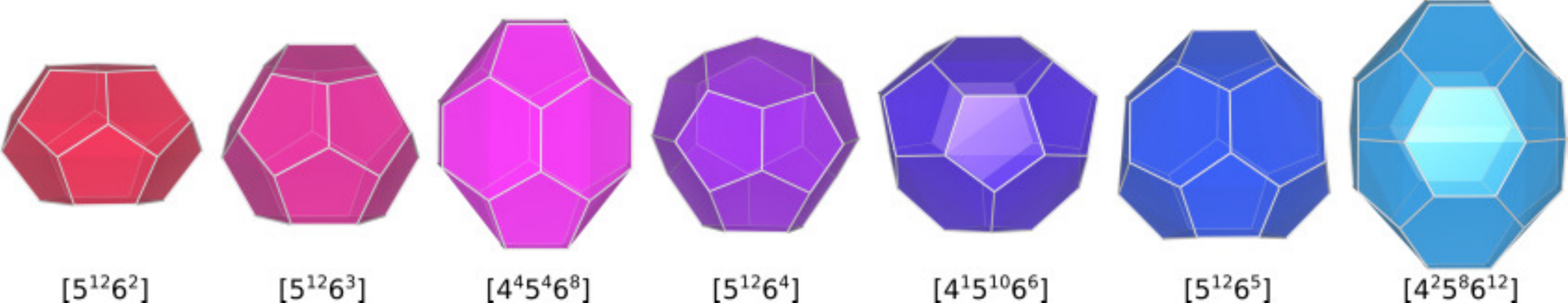}}
{\includegraphics[height=0.12\textwidth]{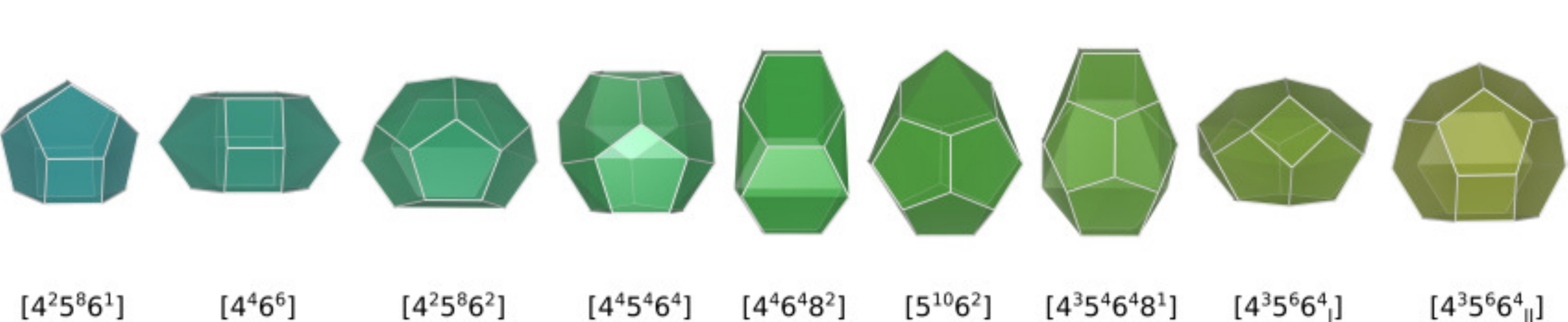}}
{\includegraphics[height=0.12\textwidth]{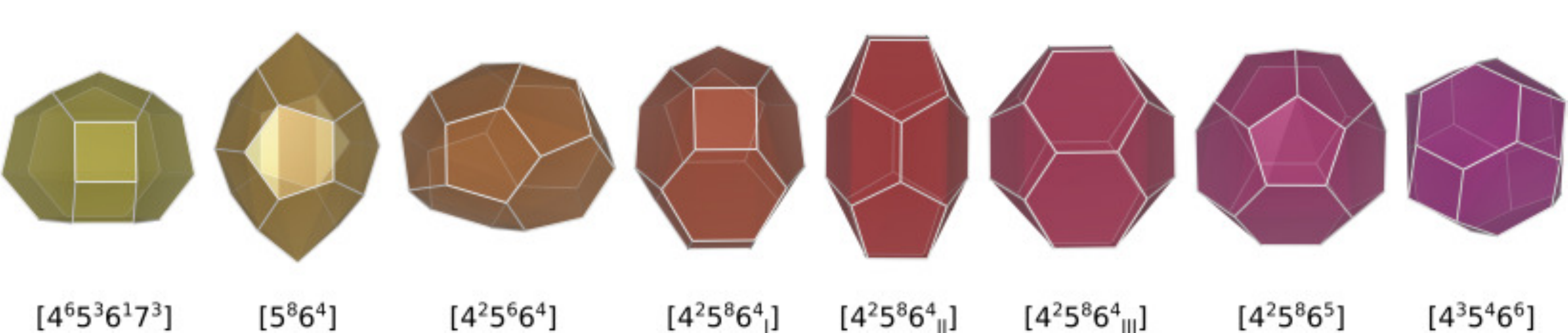}}
{\includegraphics[height=0.12\textwidth]{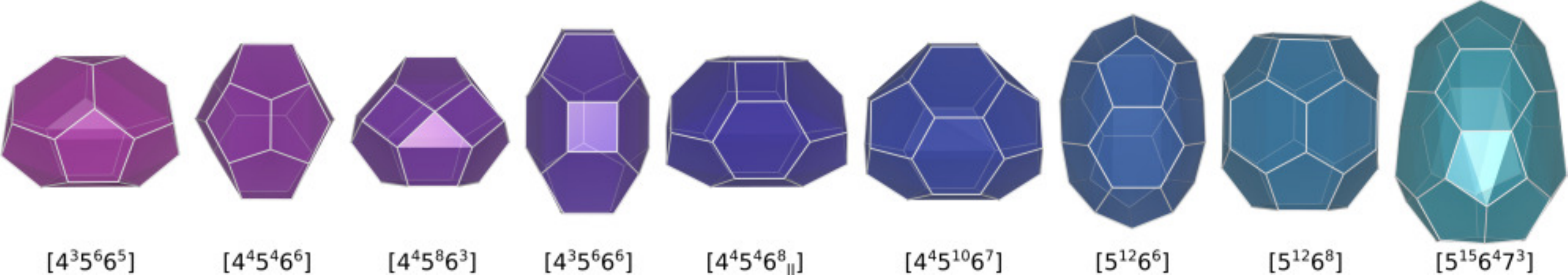}}
\label{fig:polyhedra_0}
 \end{figure*}

In contrast to the first class of structures, the second contains fully 3-dimensional cages of different types and sizes as shown in Figure~\ref{fig:polyhedra_0}. Similar to the known clathrate structures, the basic building blocks are polyhedra with a number of silicon atoms ranging from 14 to 40, consisting of faces with 4, 5, 6, 7 and 8-membered rings. Each polyhedron can be described by a set of indexes with corresponding superscripts in square brackets, $[A^\alpha B^\beta C^\gamma\dots]$, where $A$, $B$, $C$,~$\dots$ correspond to the number of vertexes of each face, and  $\alpha$, $\beta$, $\gamma$,~$\dots$ correspond to their multiplicity. Thus, the type I clathrate would be described as $[5^{12}]_2[5^{12}6^{2}]_6$, indicating that the structure is composed of 2 pentagonal dodecahedra and 6 hexagonal truncated trapezohedra. Overall, cage-like voids lead to significantly lower densities than channels in the material.

\begin{figure*}[h!]            
\caption{Low-density silicon allotropes formed of cage-like building blocks. The colors of the polyhedra correspond to the ones in Figure~\ref{fig:polyhedra_0}.}
 \centering
\setlength{\unitlength}{1cm}

\hfill{\includegraphics[height=0.16\textwidth]{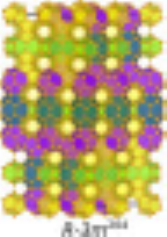}} 
\hfill{\includegraphics[height=0.16\textwidth]{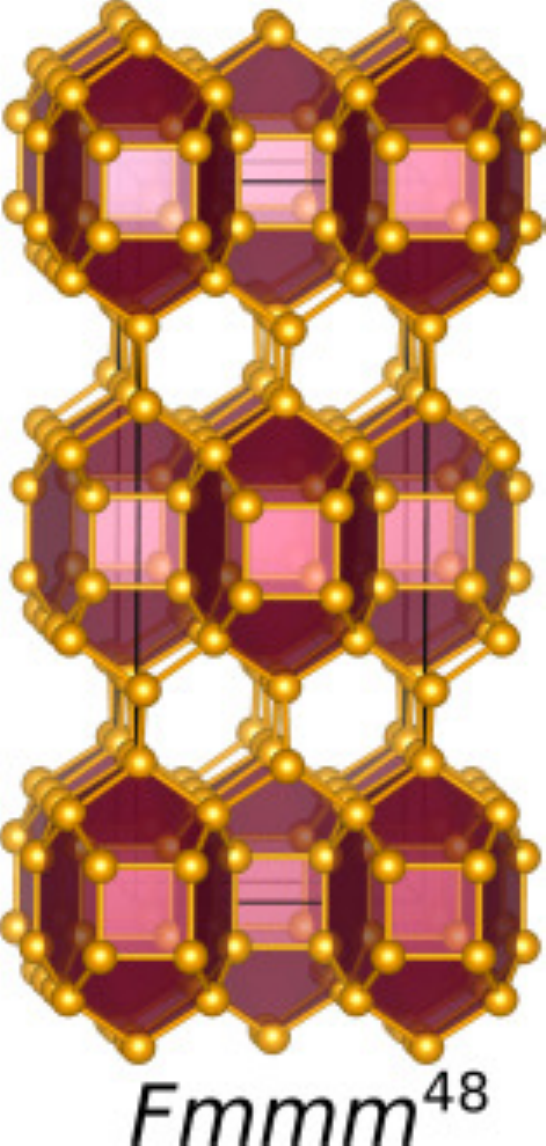}} 
\hfill{\includegraphics[height=0.16\textwidth]{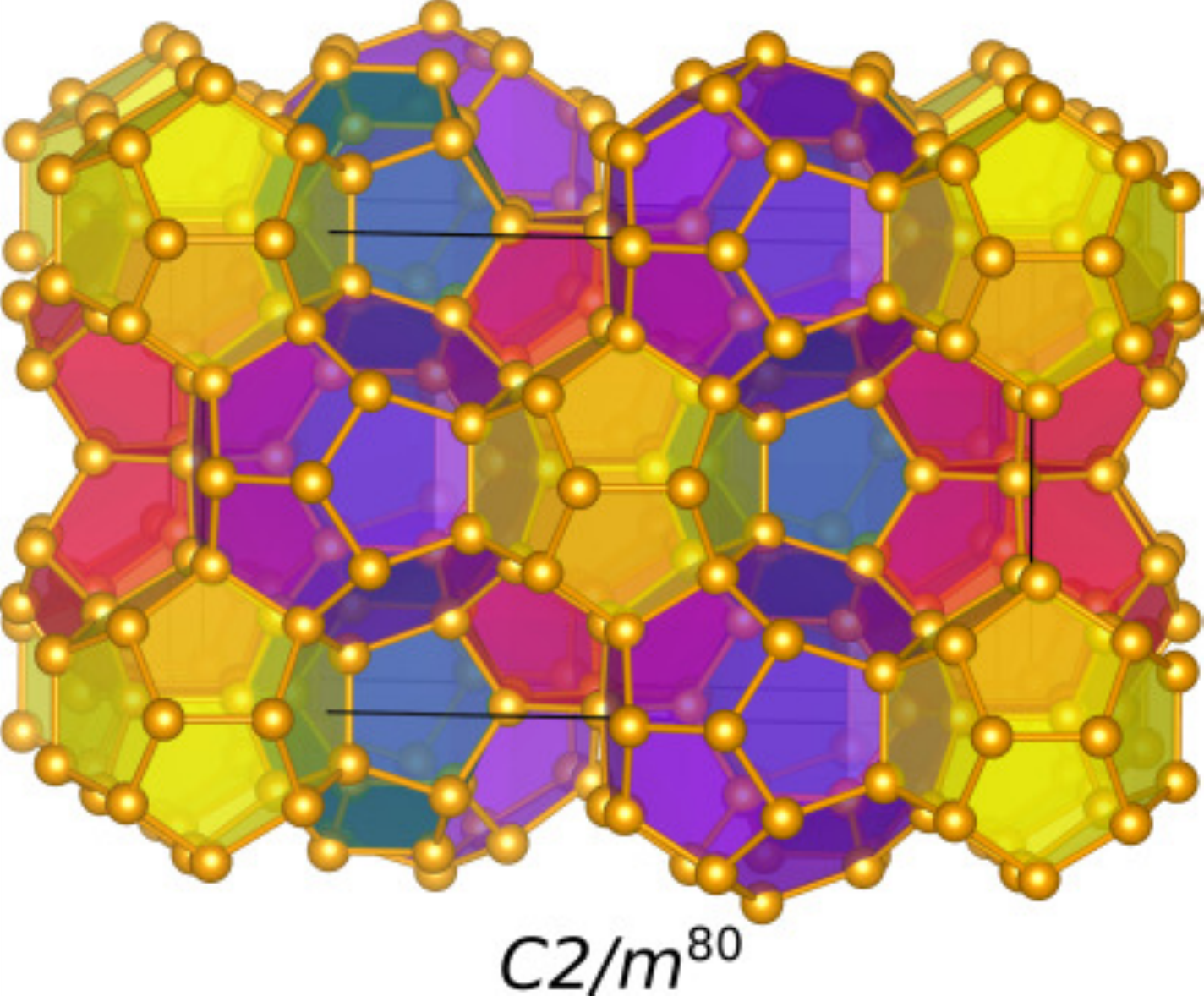}} 
\hfill{\includegraphics[height=0.16\textwidth]{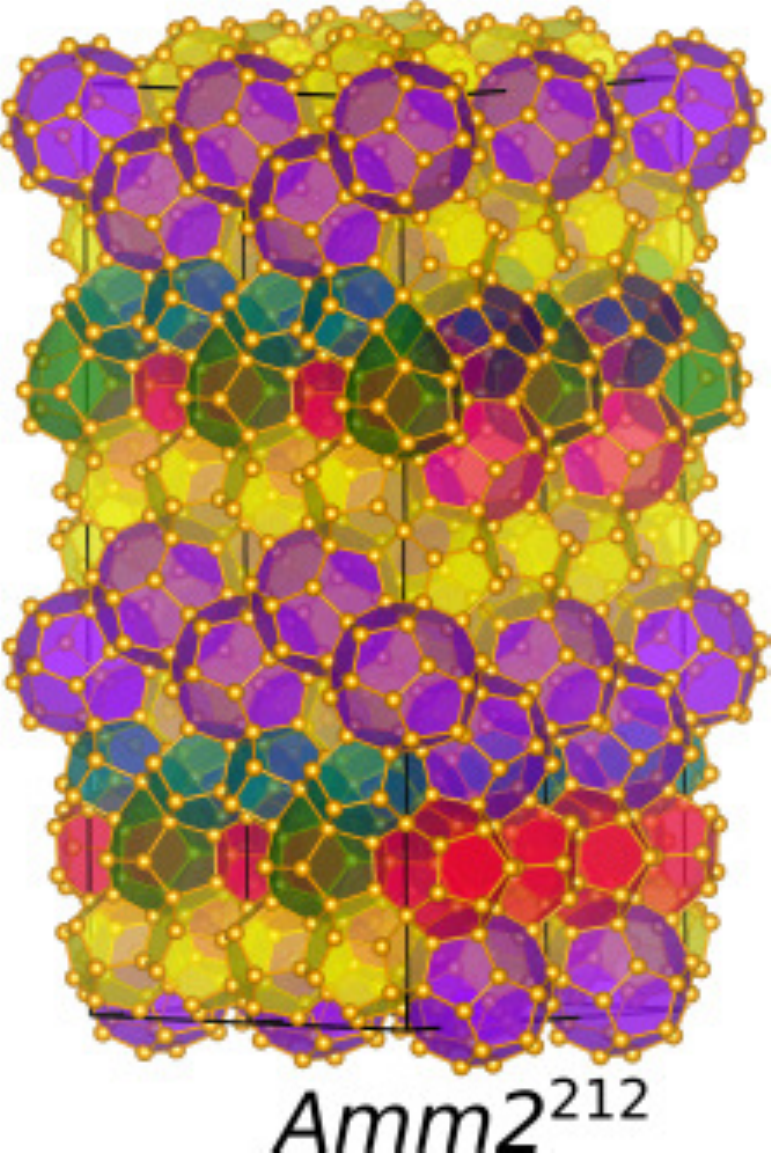}} 
\hfill{\includegraphics[height=0.16\textwidth]{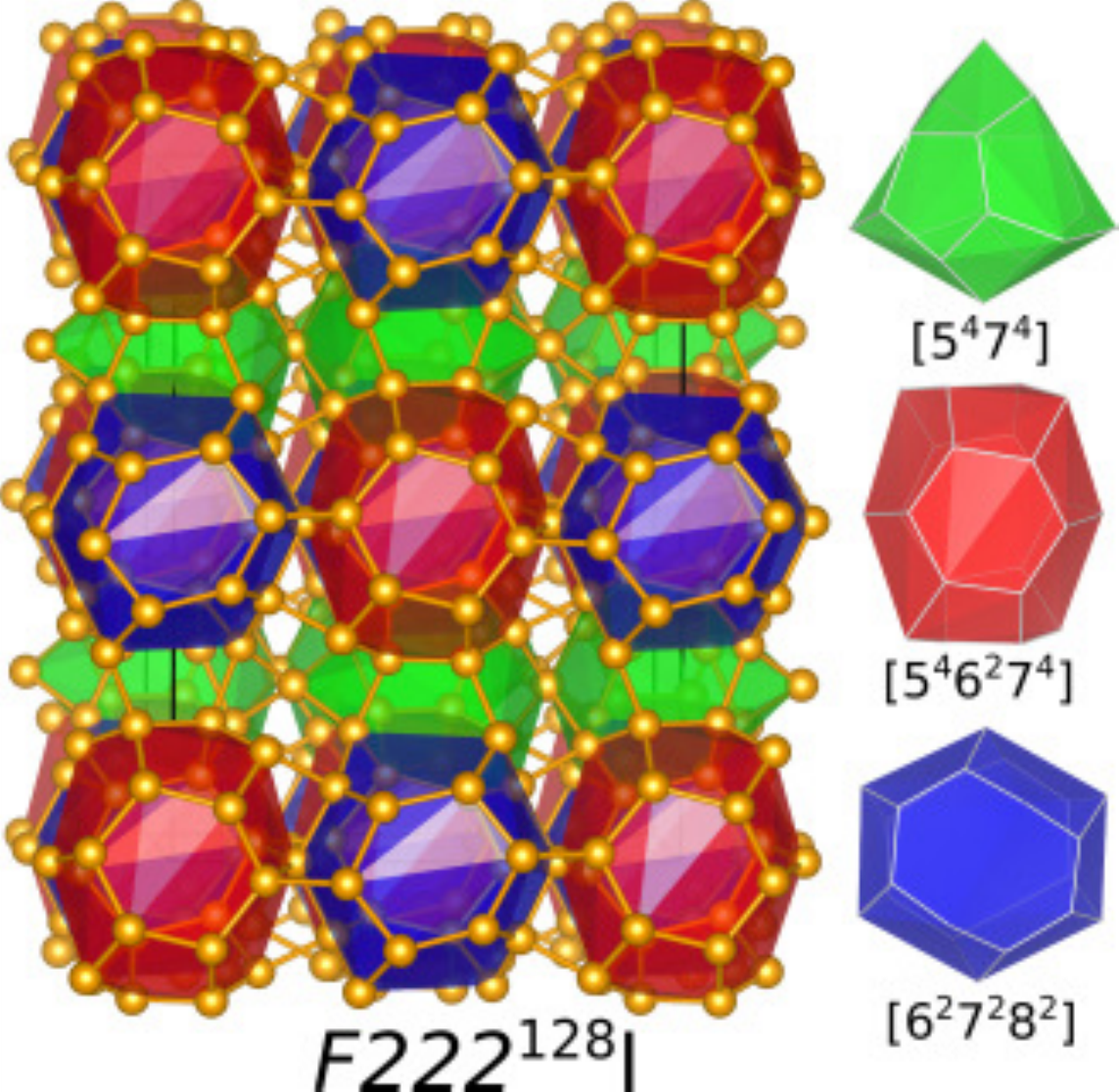}} 
\hfill{\includegraphics[height=0.16\textwidth]{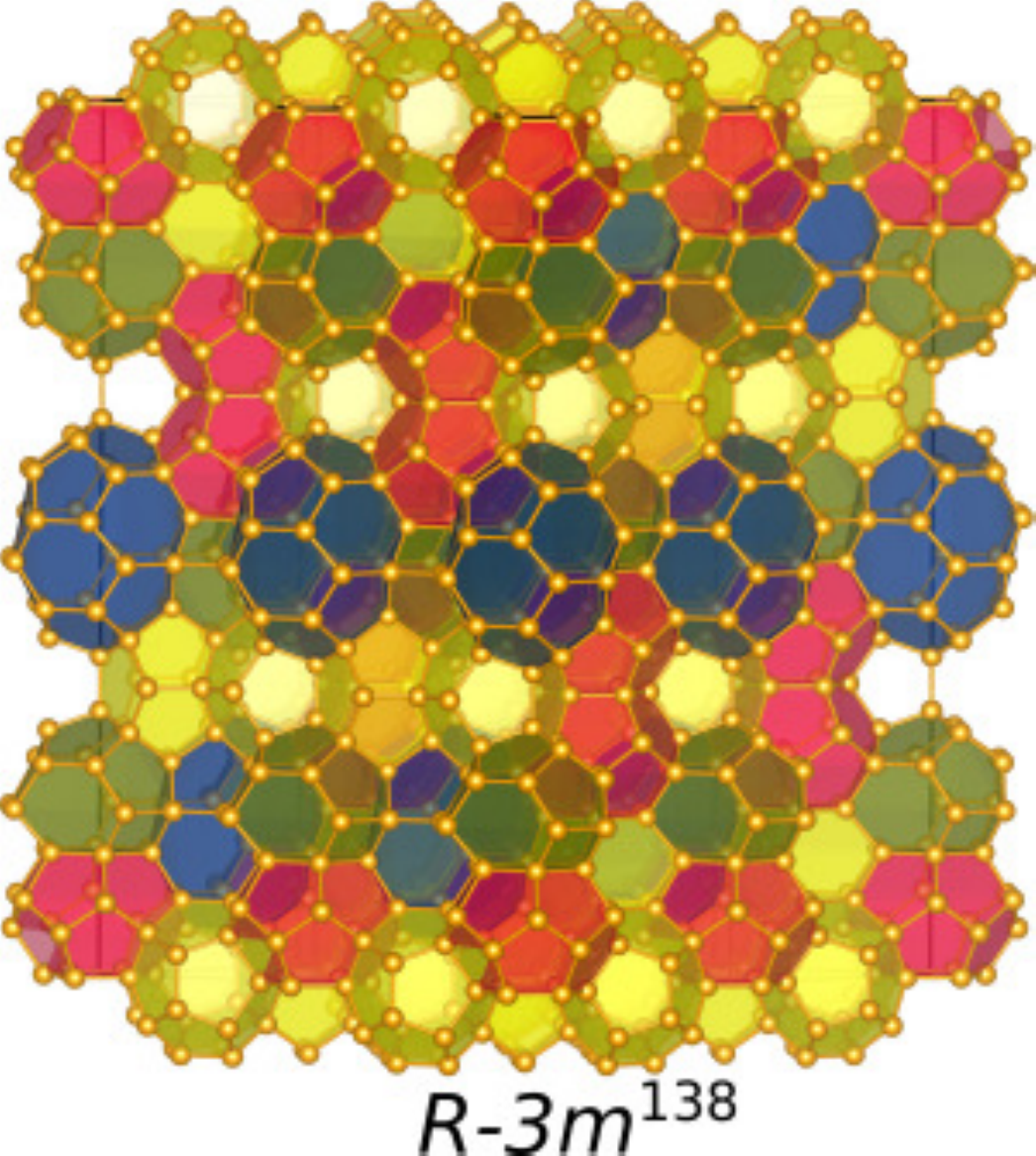}} 
\hfill{\includegraphics[height=0.16\textwidth]{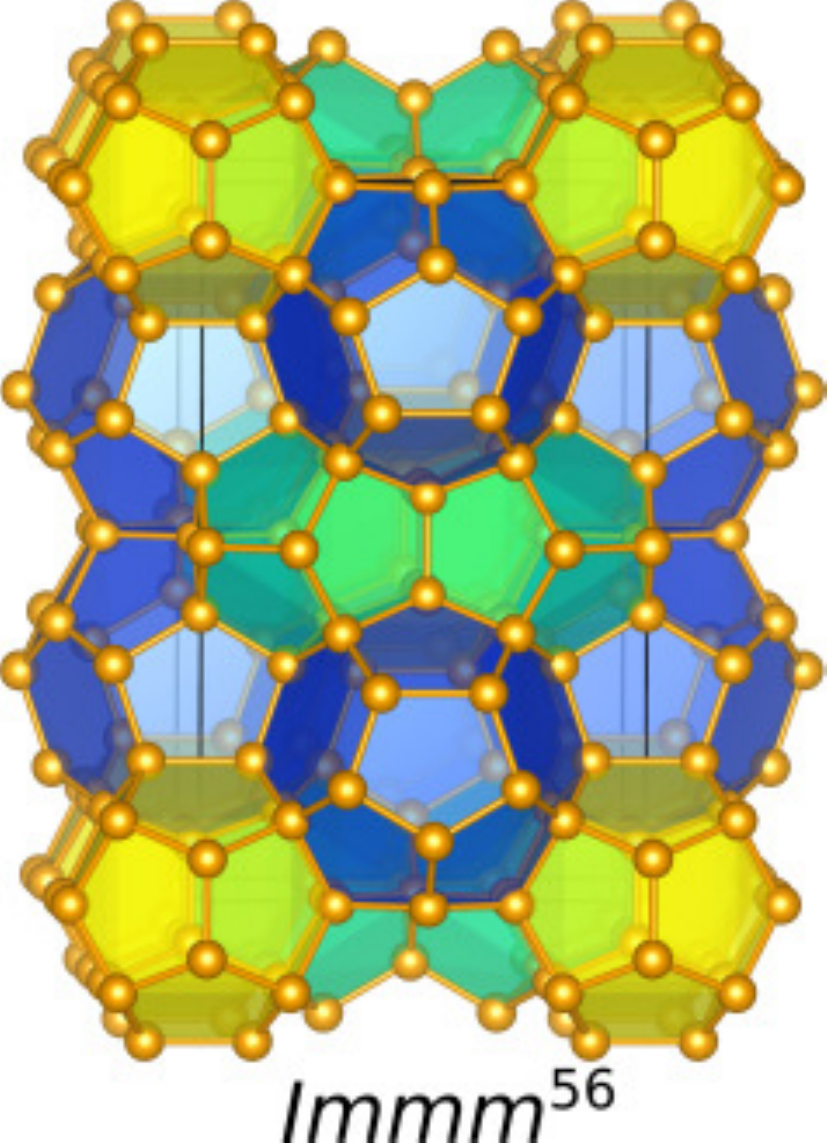}}
\hfill{\includegraphics[height=0.16\textwidth]{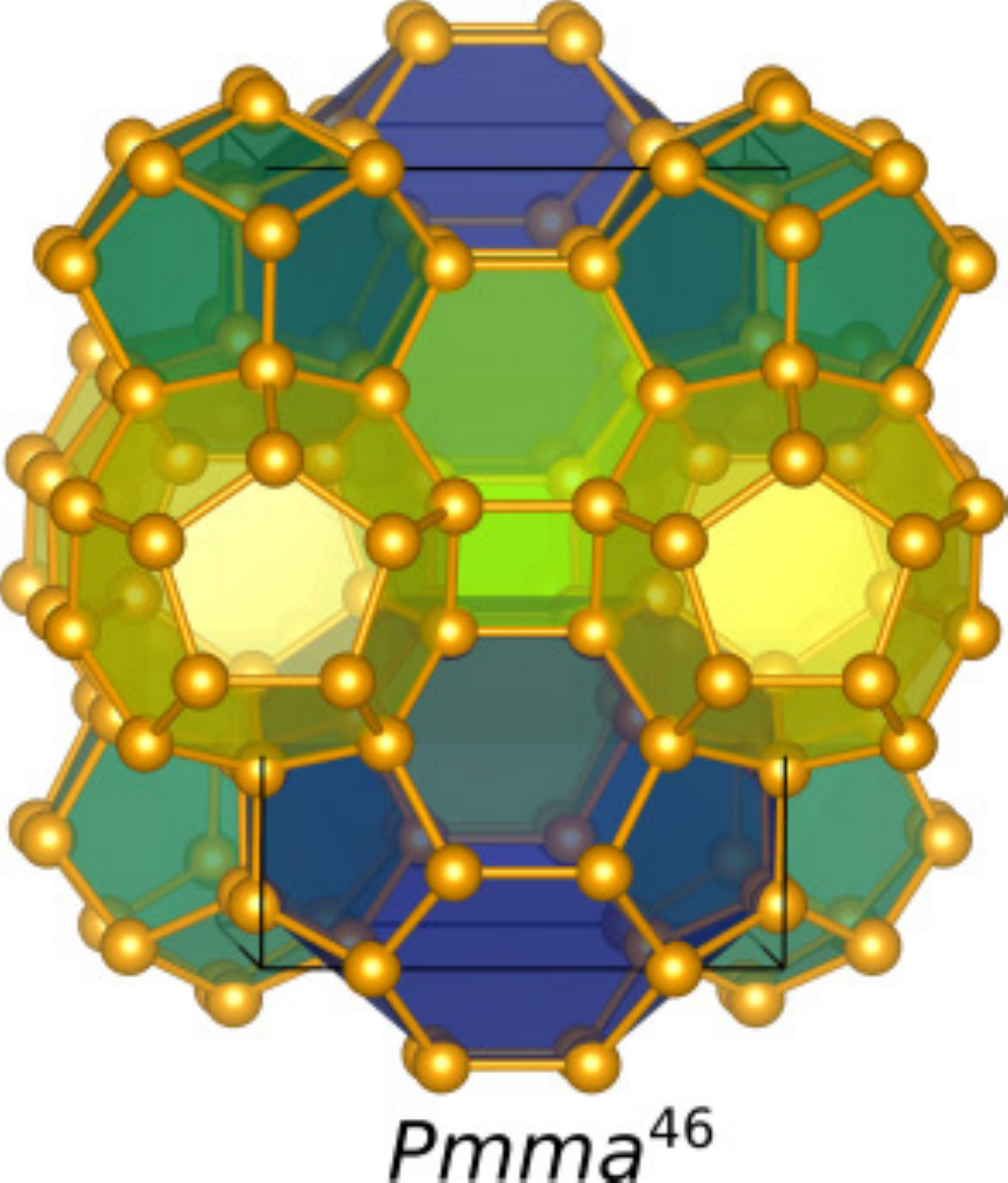}}
\hfill{\includegraphics[height=0.16\textwidth]{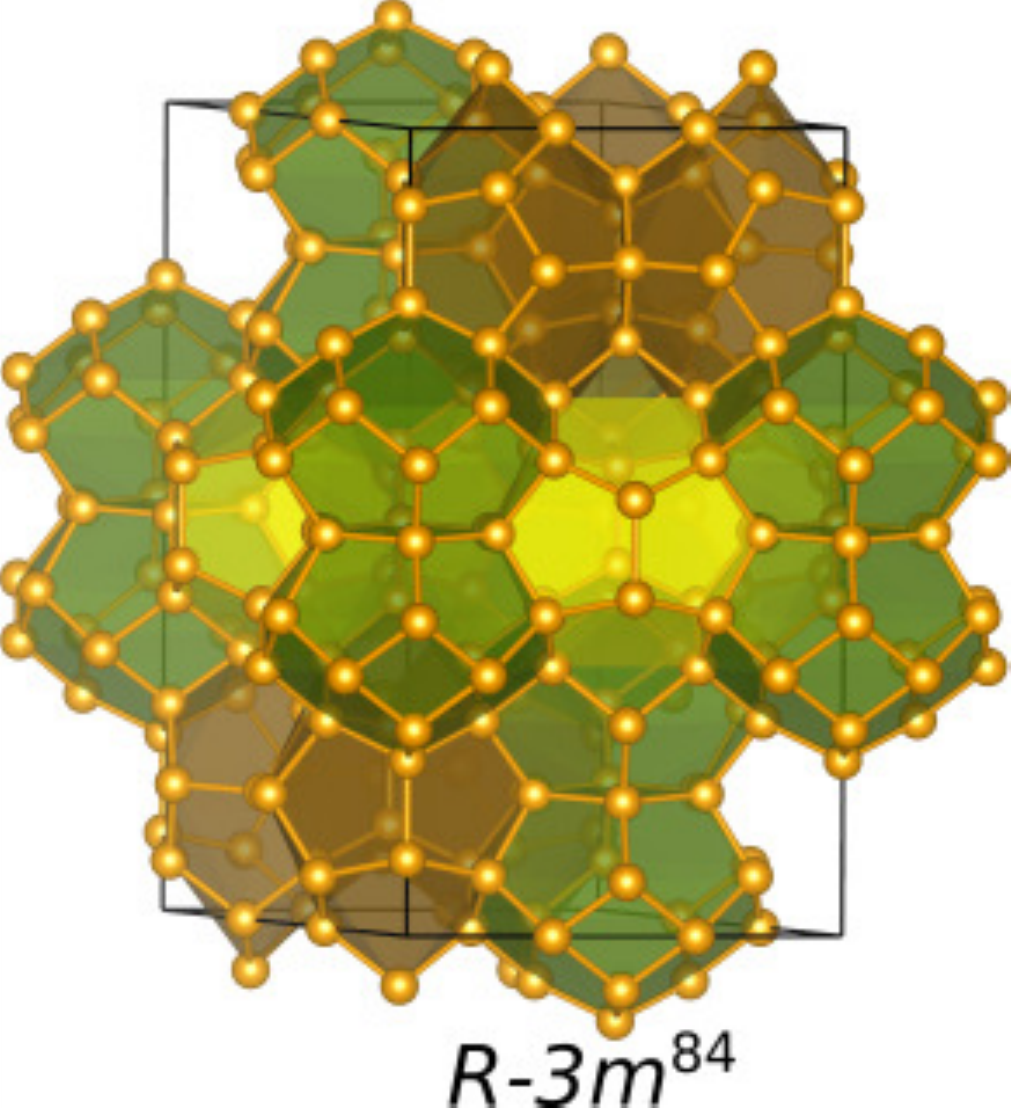}}
\hfill{\includegraphics[height=0.16\textwidth]{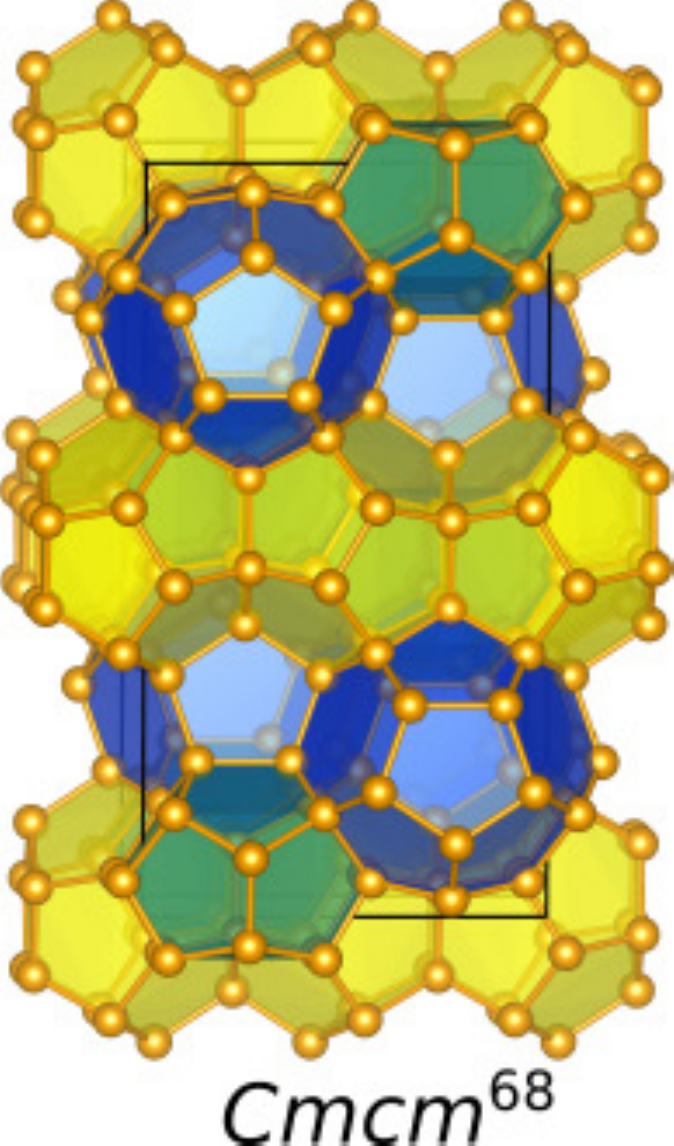}}
\hfill{\includegraphics[height=0.16\textwidth]{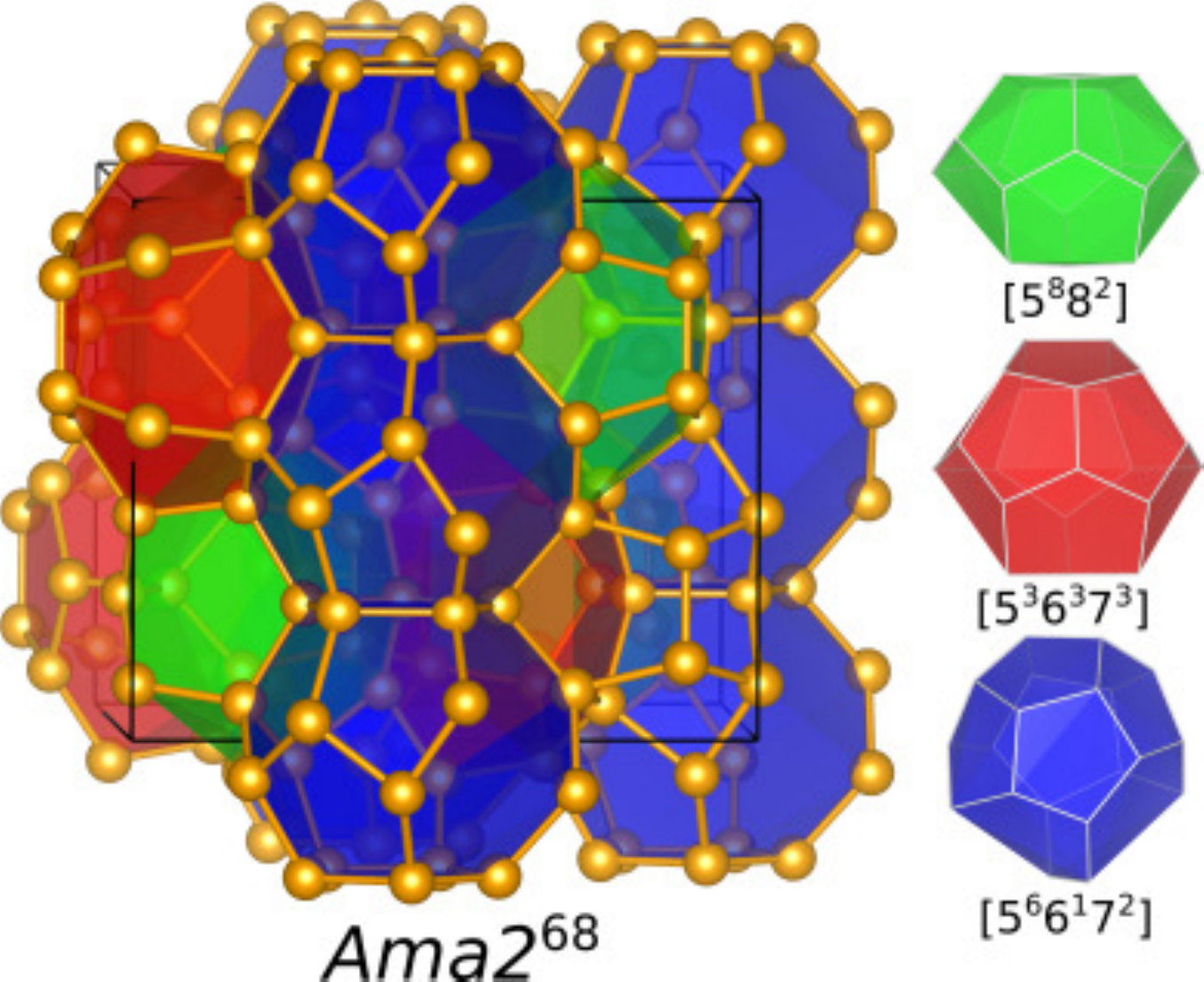}}
\hfill{\includegraphics[height=0.16\textwidth]{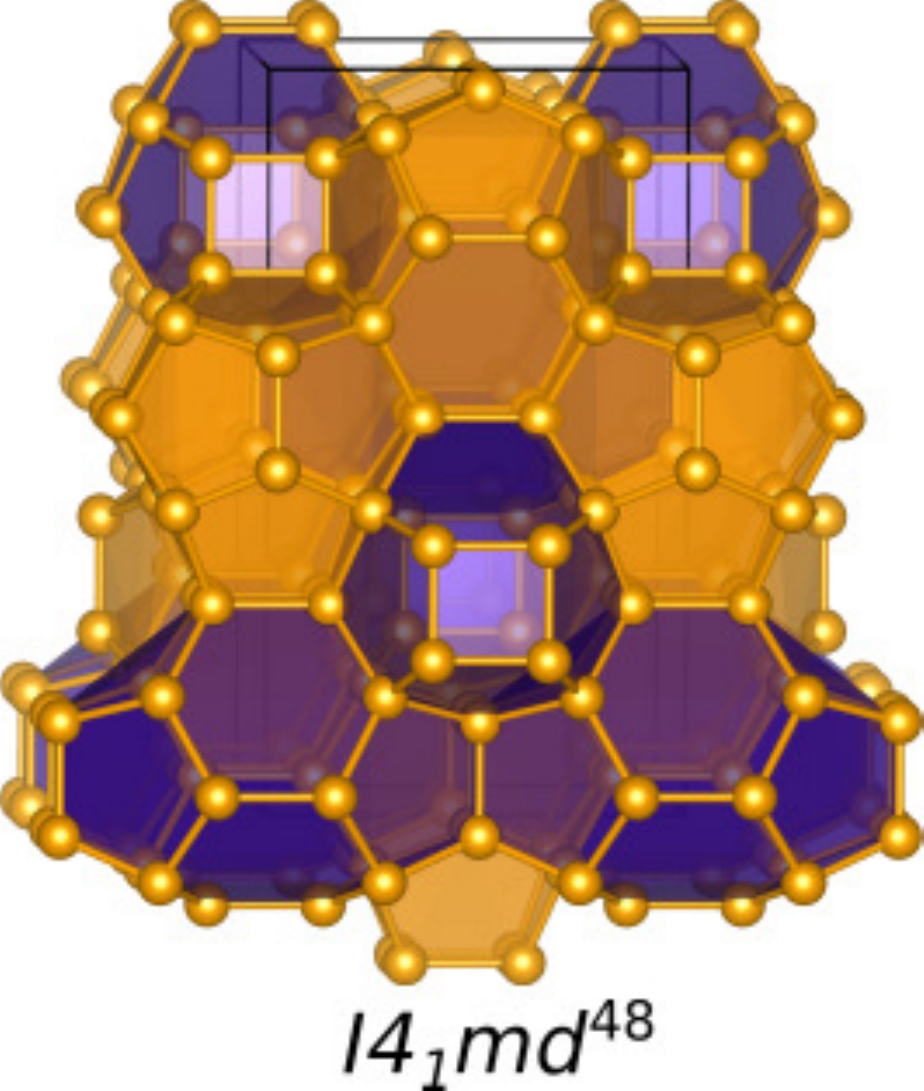}}
\hfill{\includegraphics[height=0.16\textwidth]{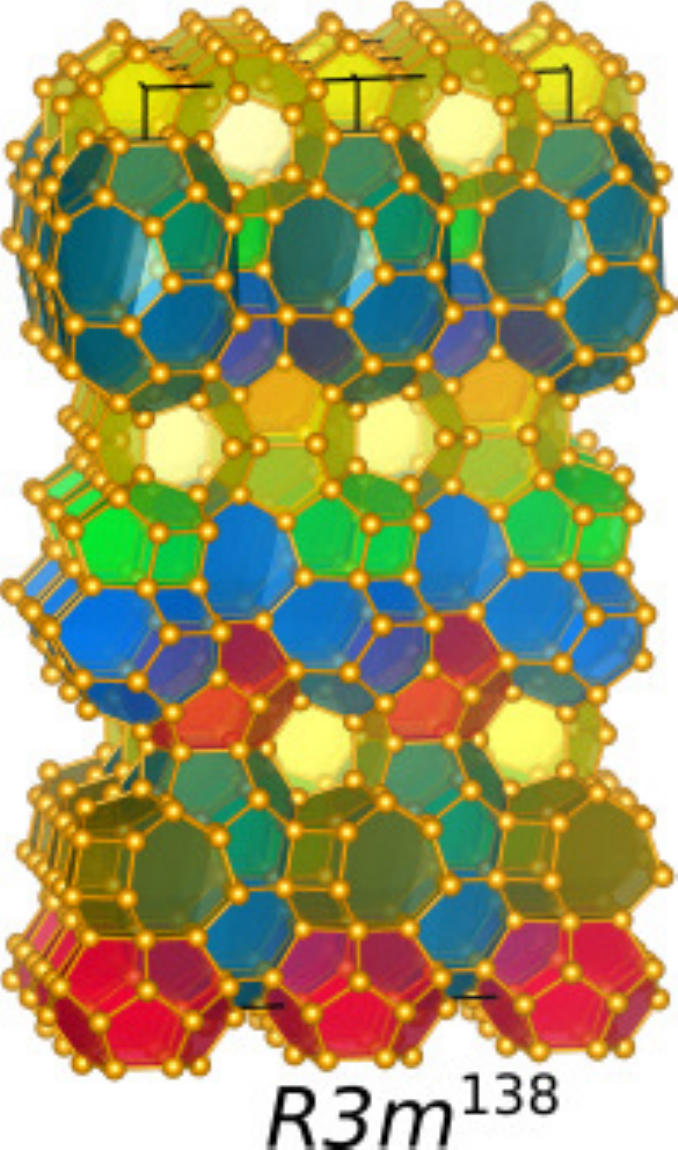}}
\hfill{\includegraphics[height=0.16\textwidth]{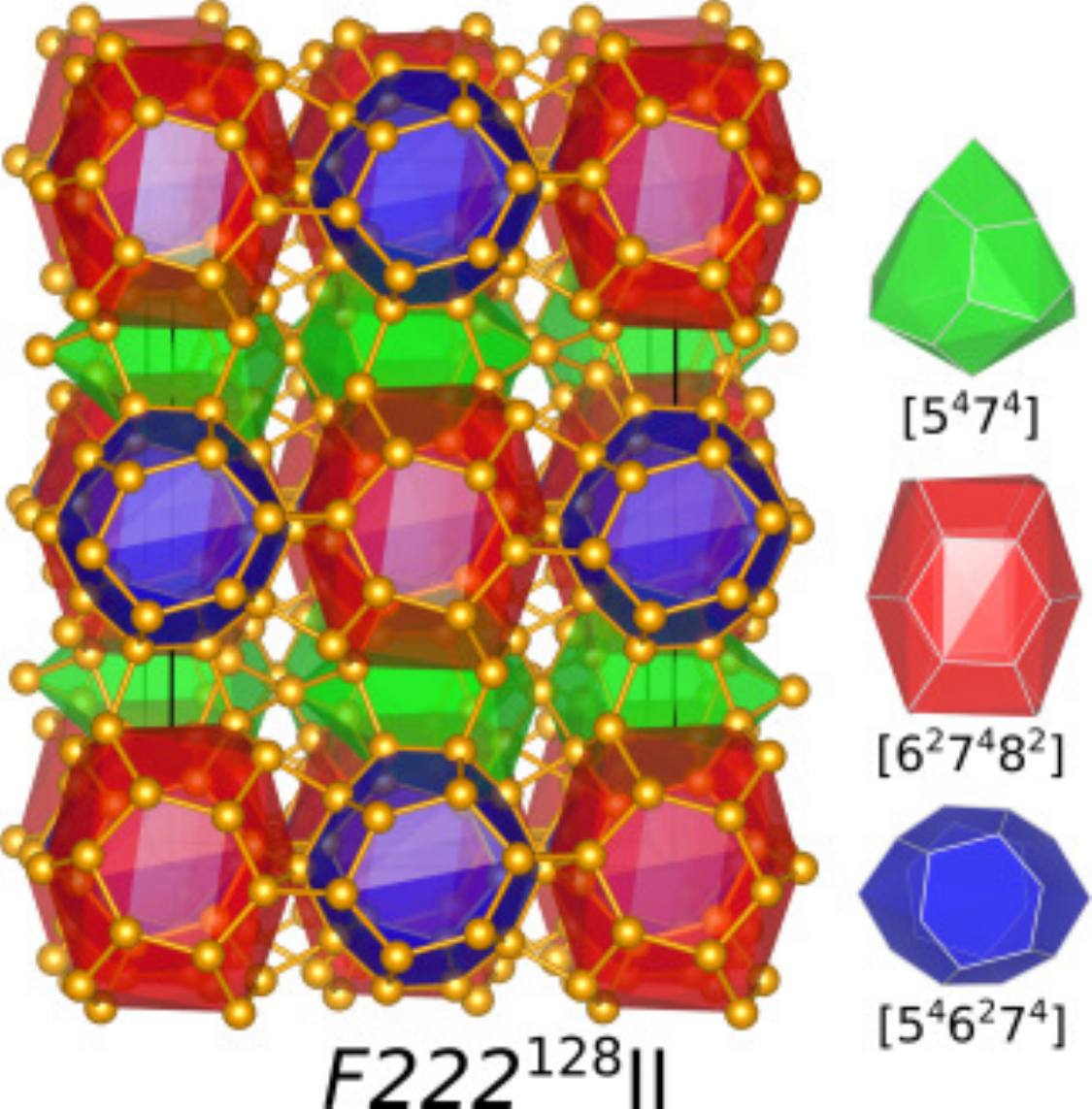}}
\hfill{\includegraphics[height=0.16\textwidth]{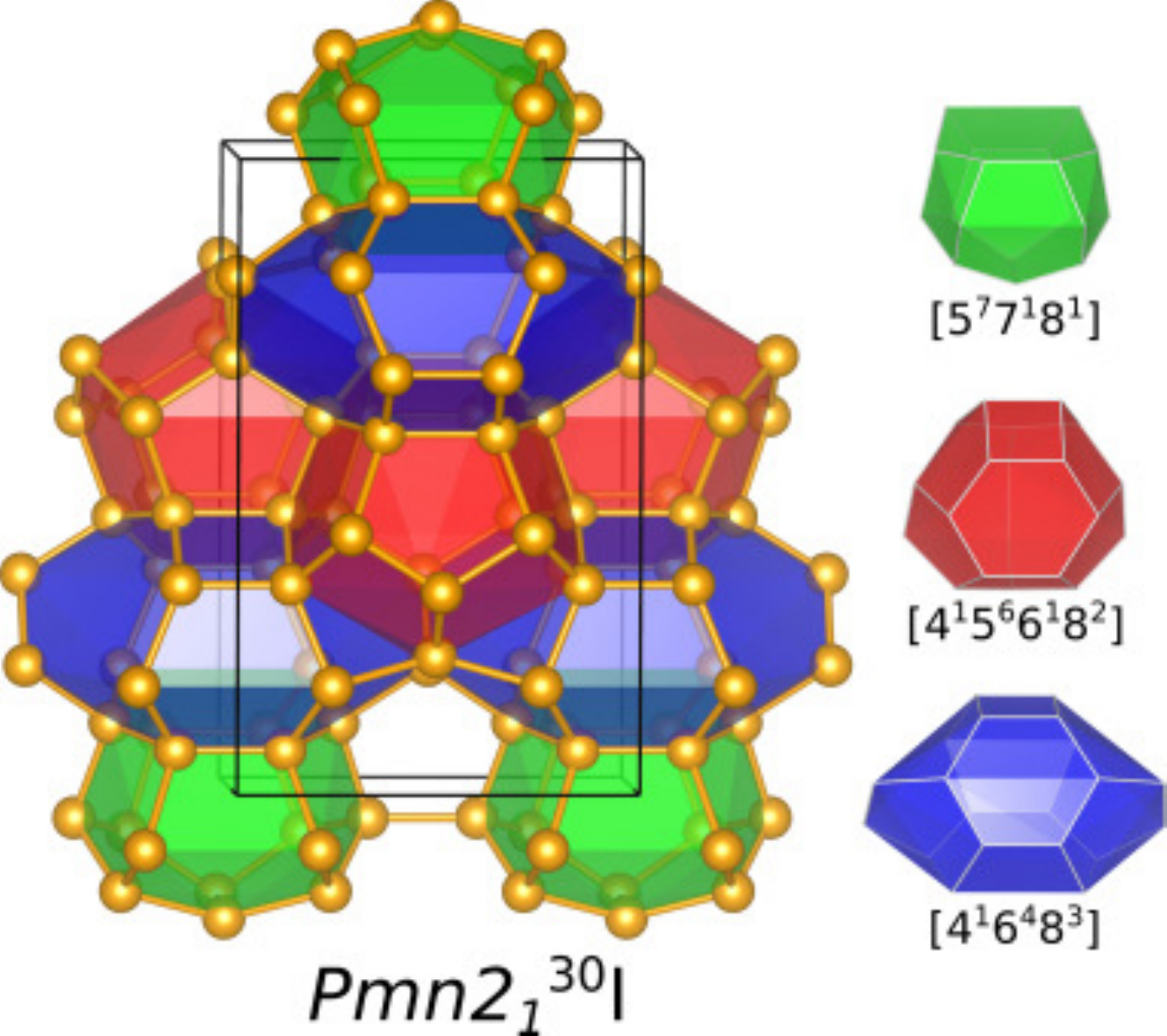}}
\hfill{\includegraphics[height=0.16\textwidth]{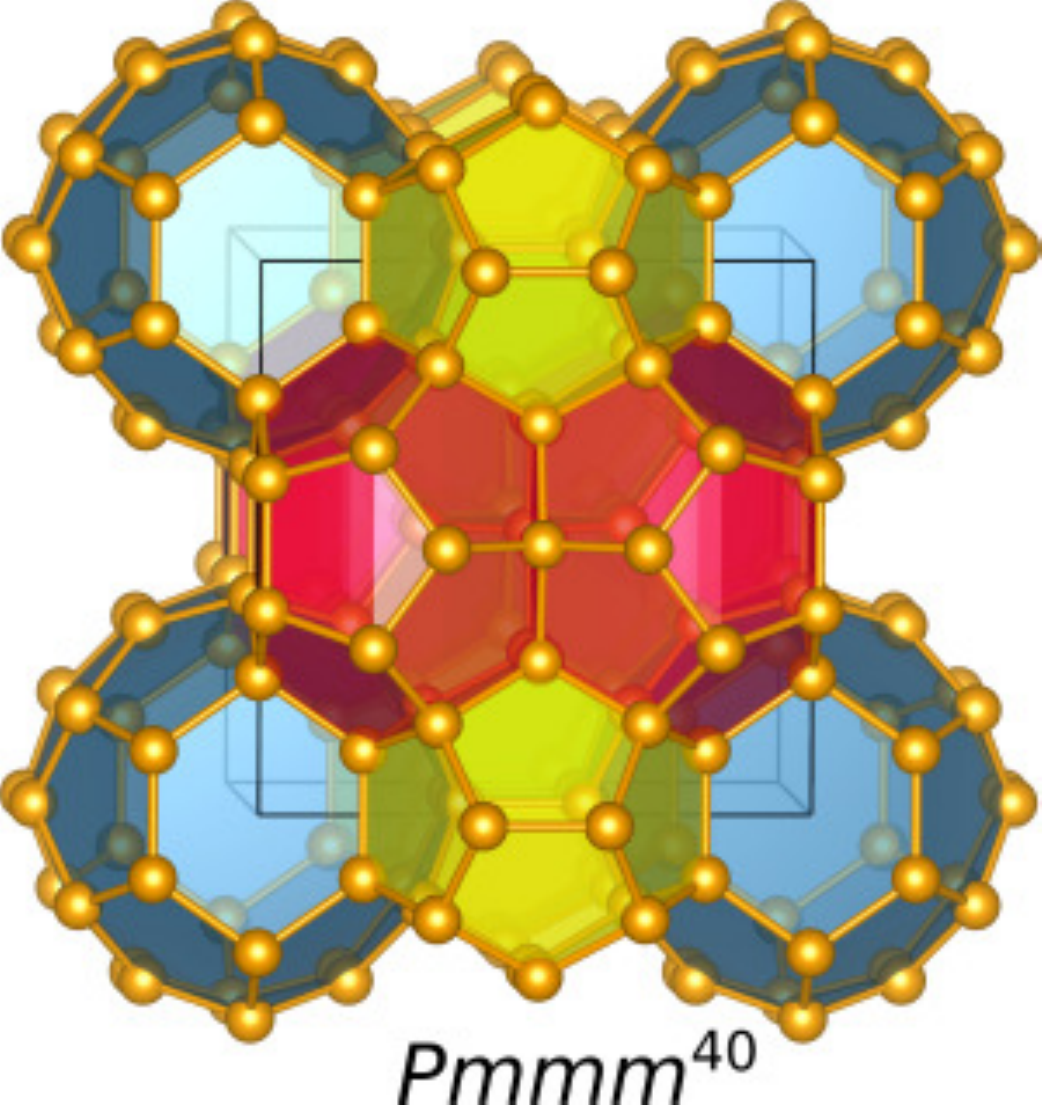}}
\hfill{\includegraphics[height=0.16\textwidth]{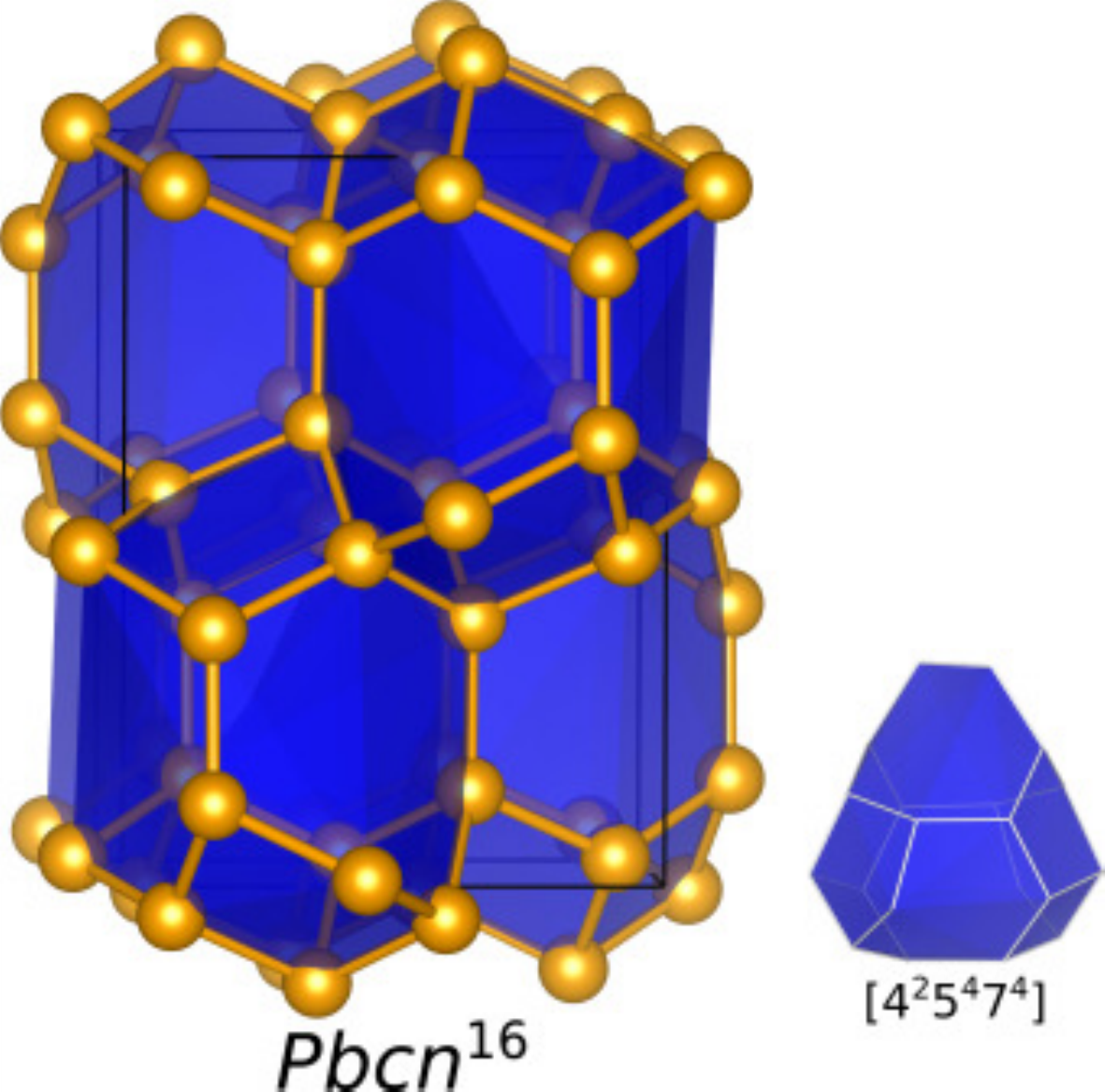}}
\hfill{\includegraphics[height=0.16\textwidth]{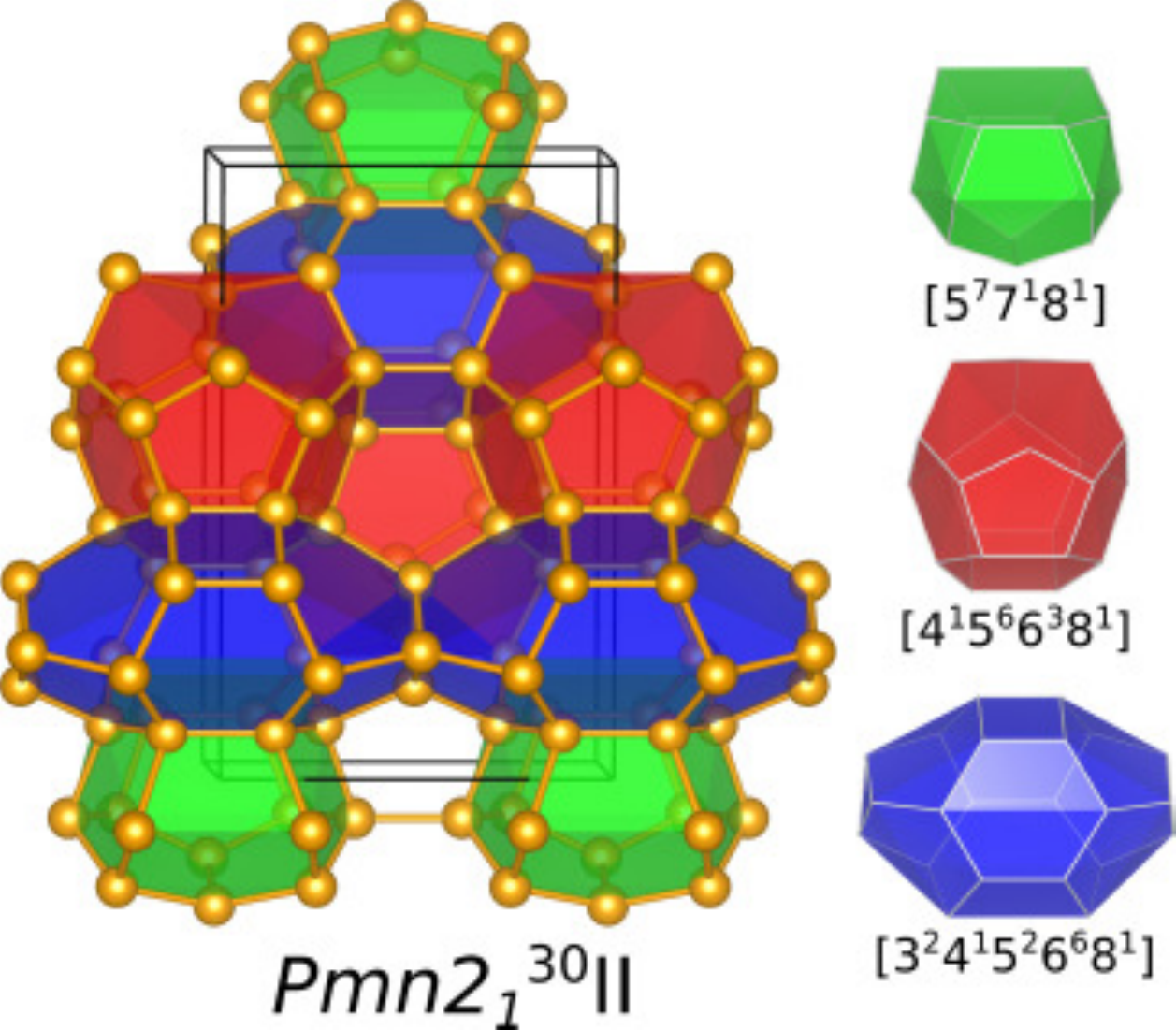}}
\hfill{\includegraphics[height=0.16\textwidth]{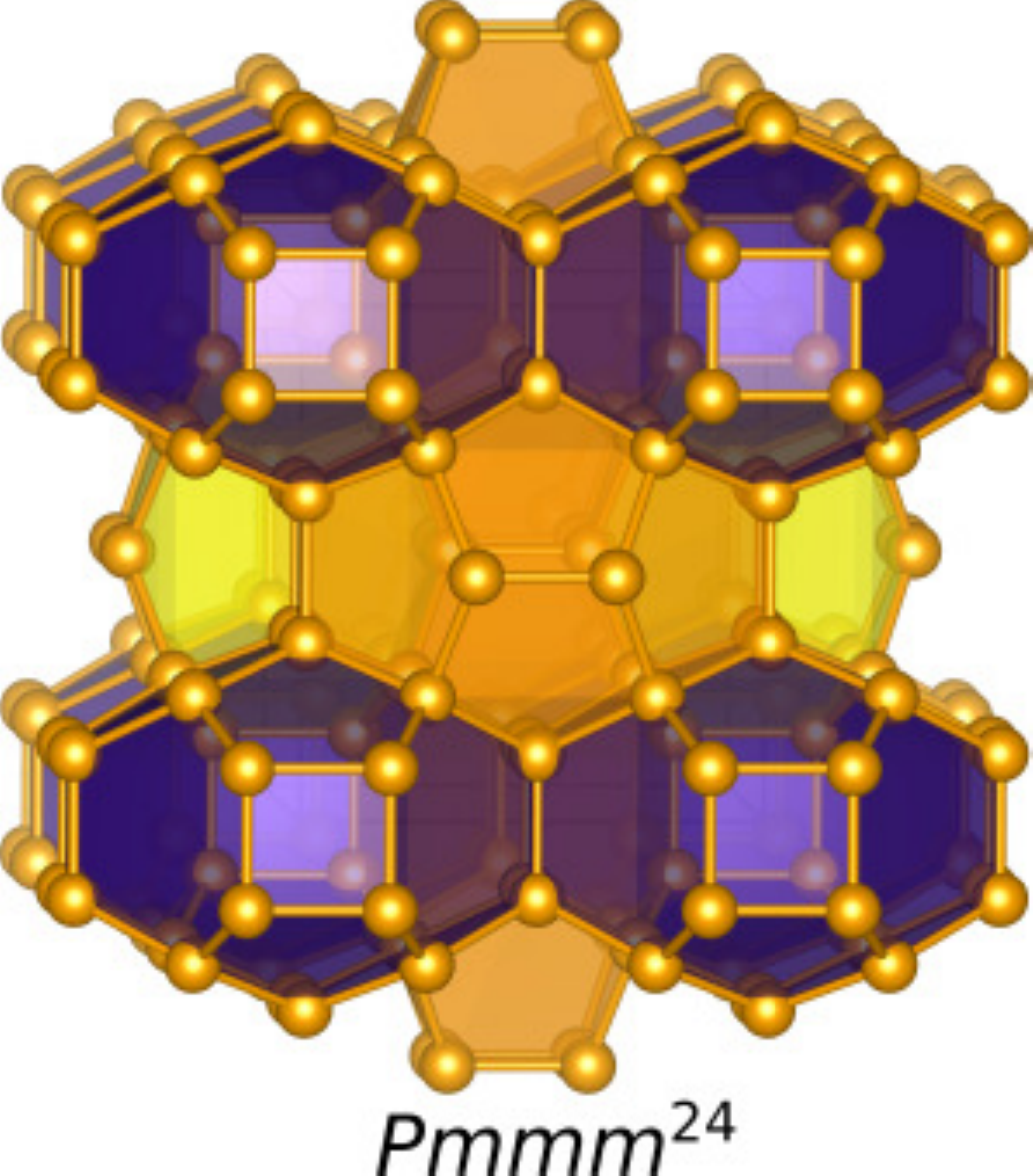}}
\hfill{\includegraphics[height=0.16\textwidth]{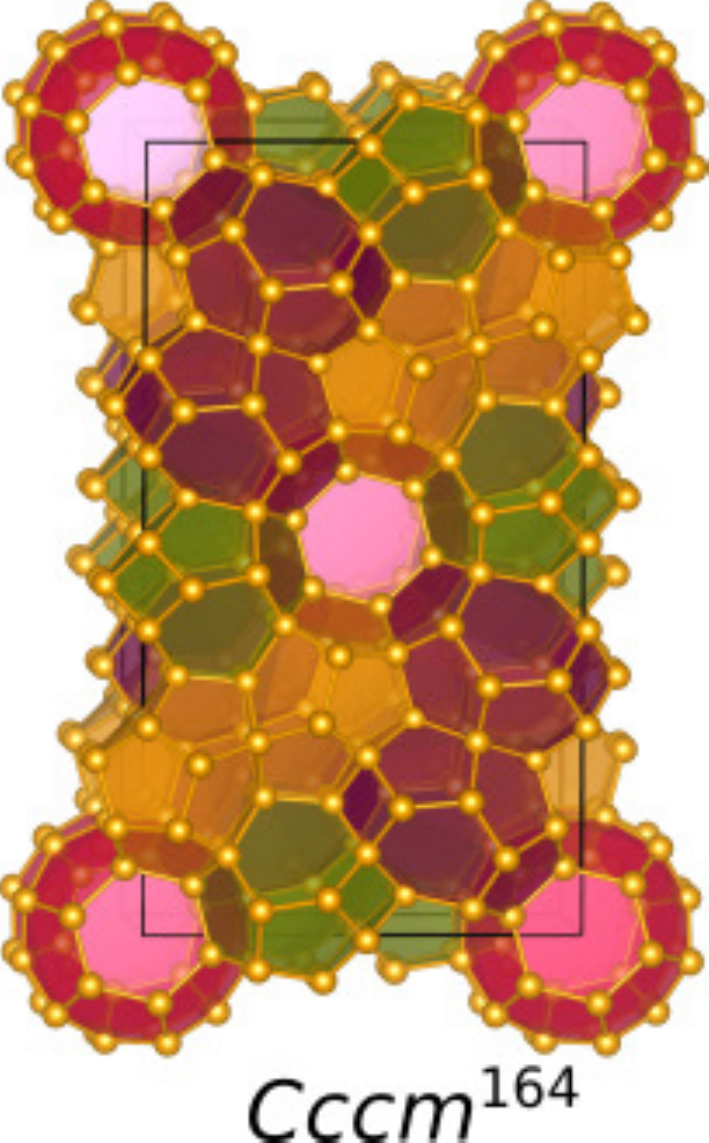}}
\hfill{\includegraphics[height=0.16\textwidth]{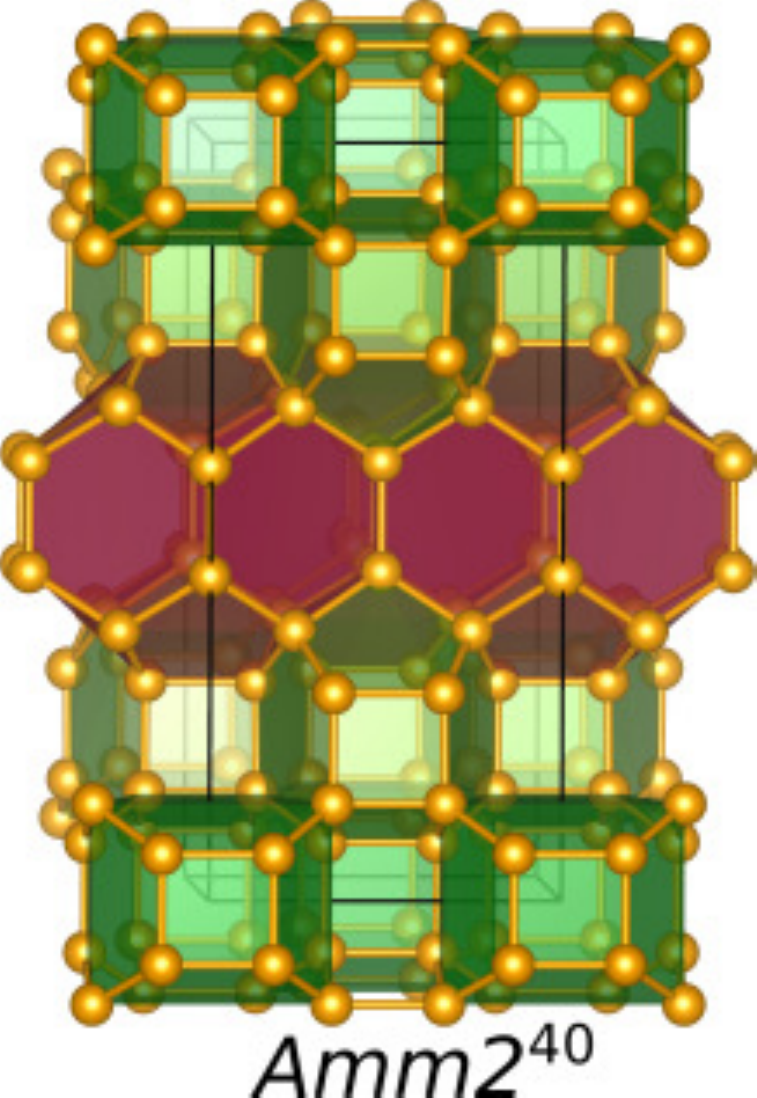}}
\hfill{\includegraphics[height=0.16\textwidth]{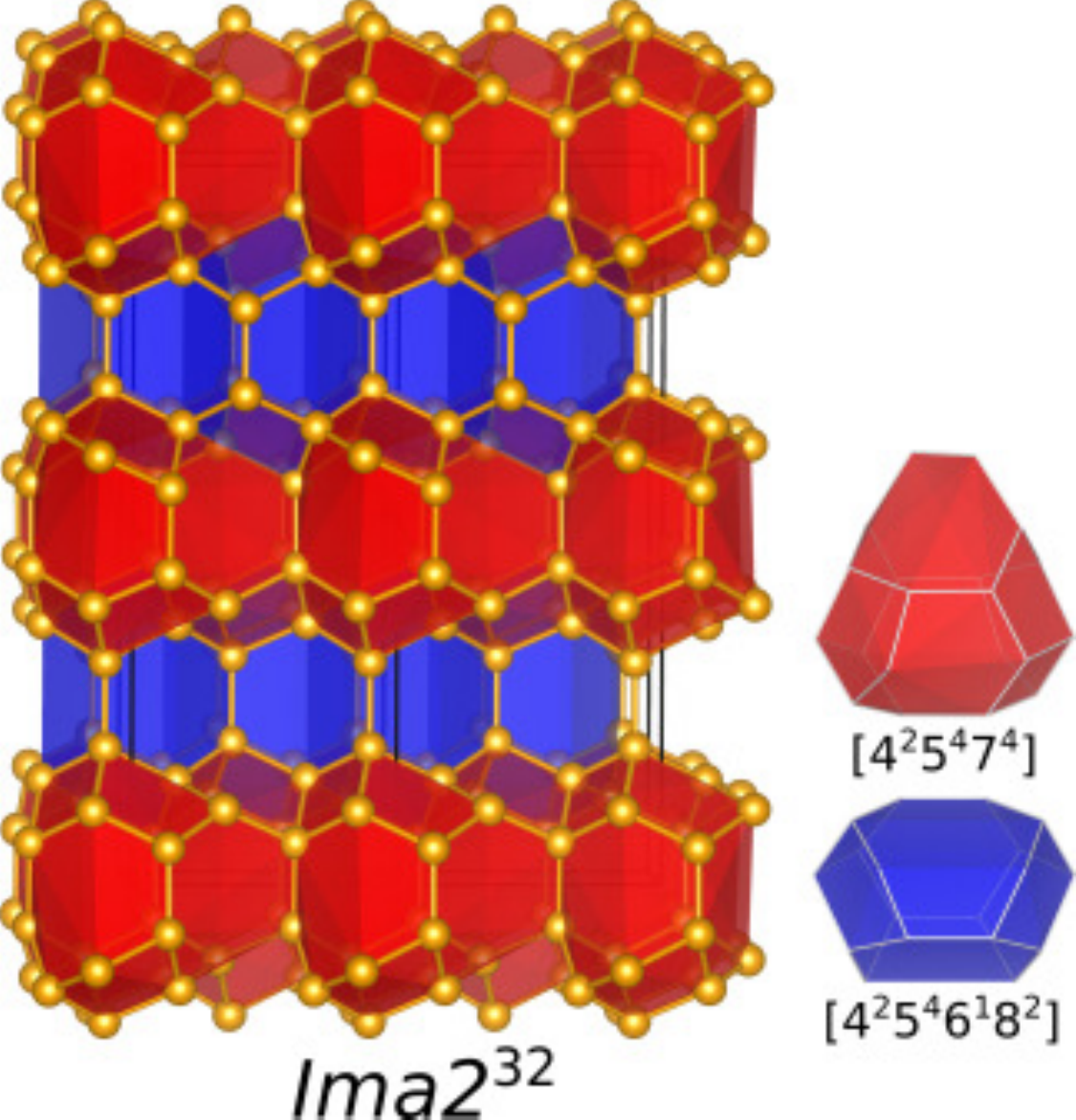}}
\hfill{\includegraphics[height=0.16\textwidth]{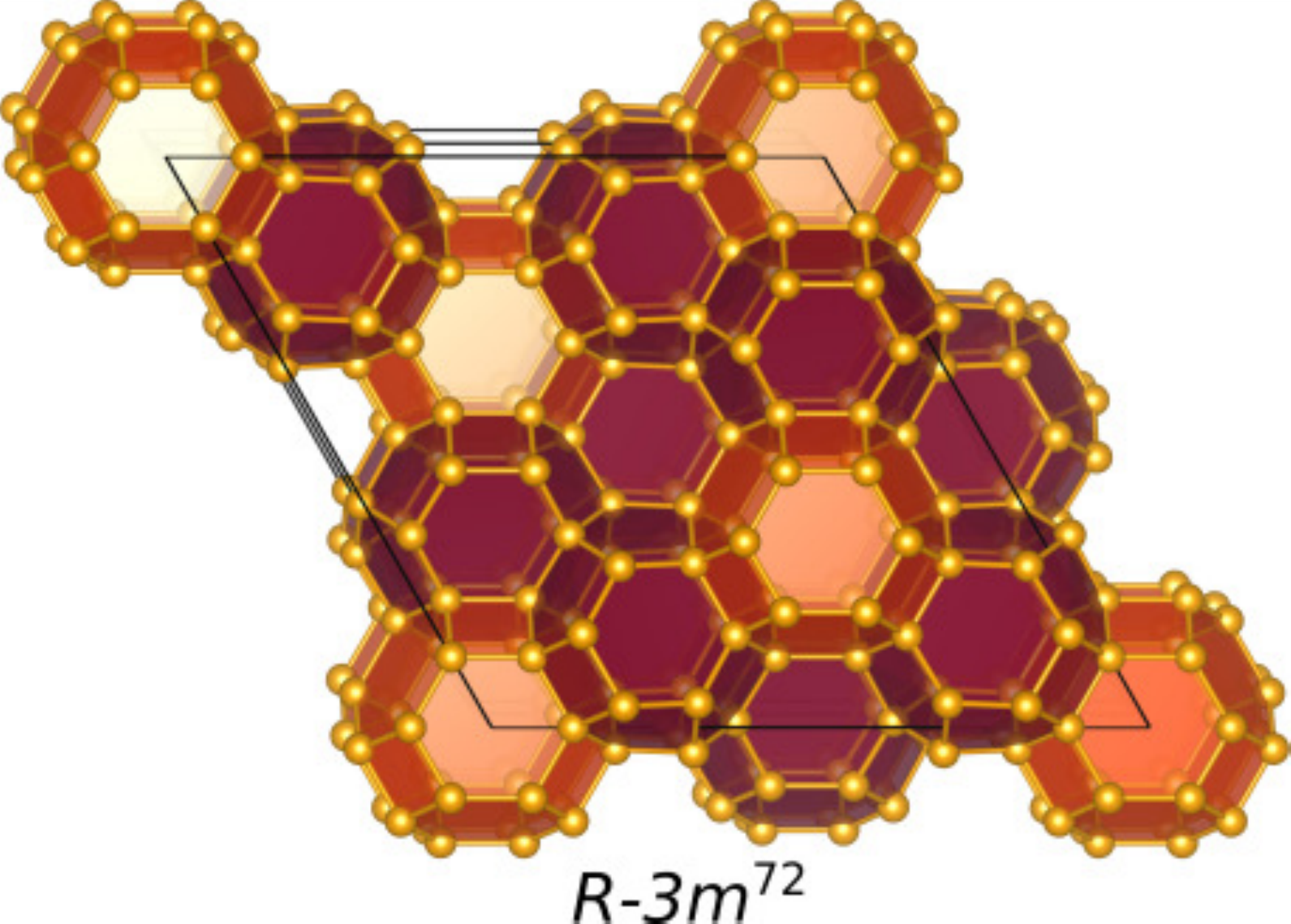}}
\hfill{\includegraphics[height=0.16\textwidth]{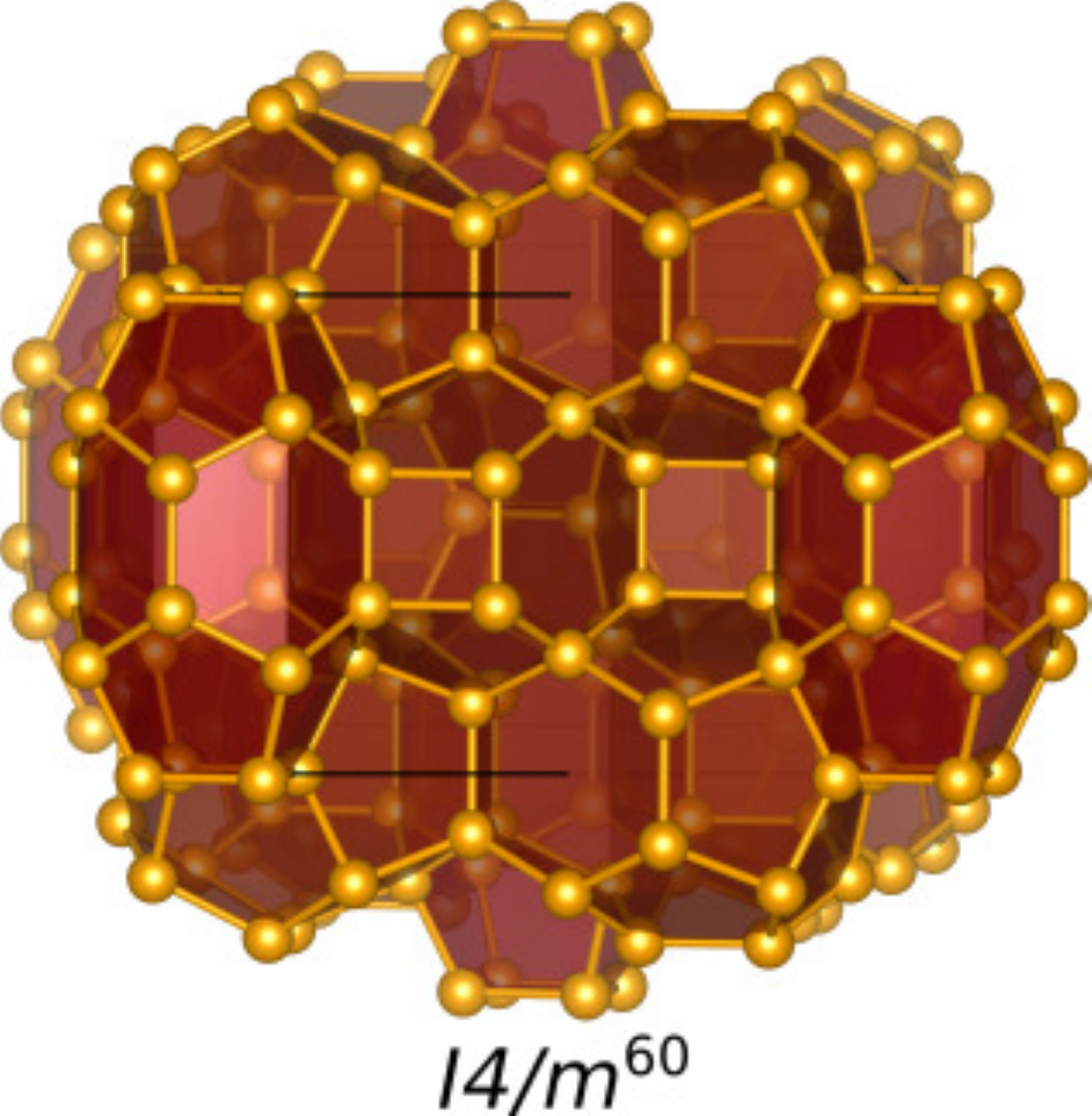}}
\hfill{\includegraphics[height=0.16\textwidth]{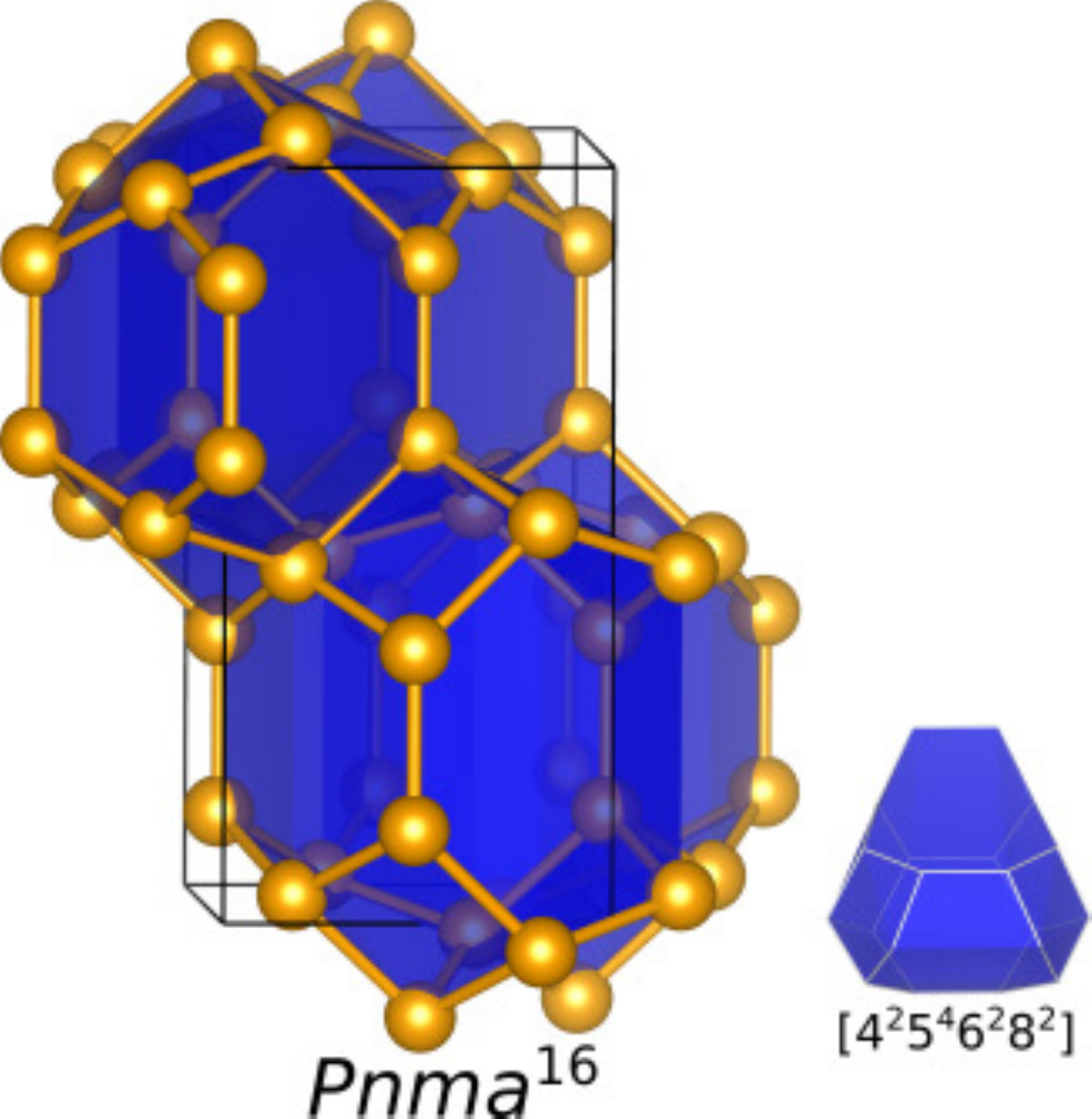}}
\hfill{\includegraphics[height=0.16\textwidth]{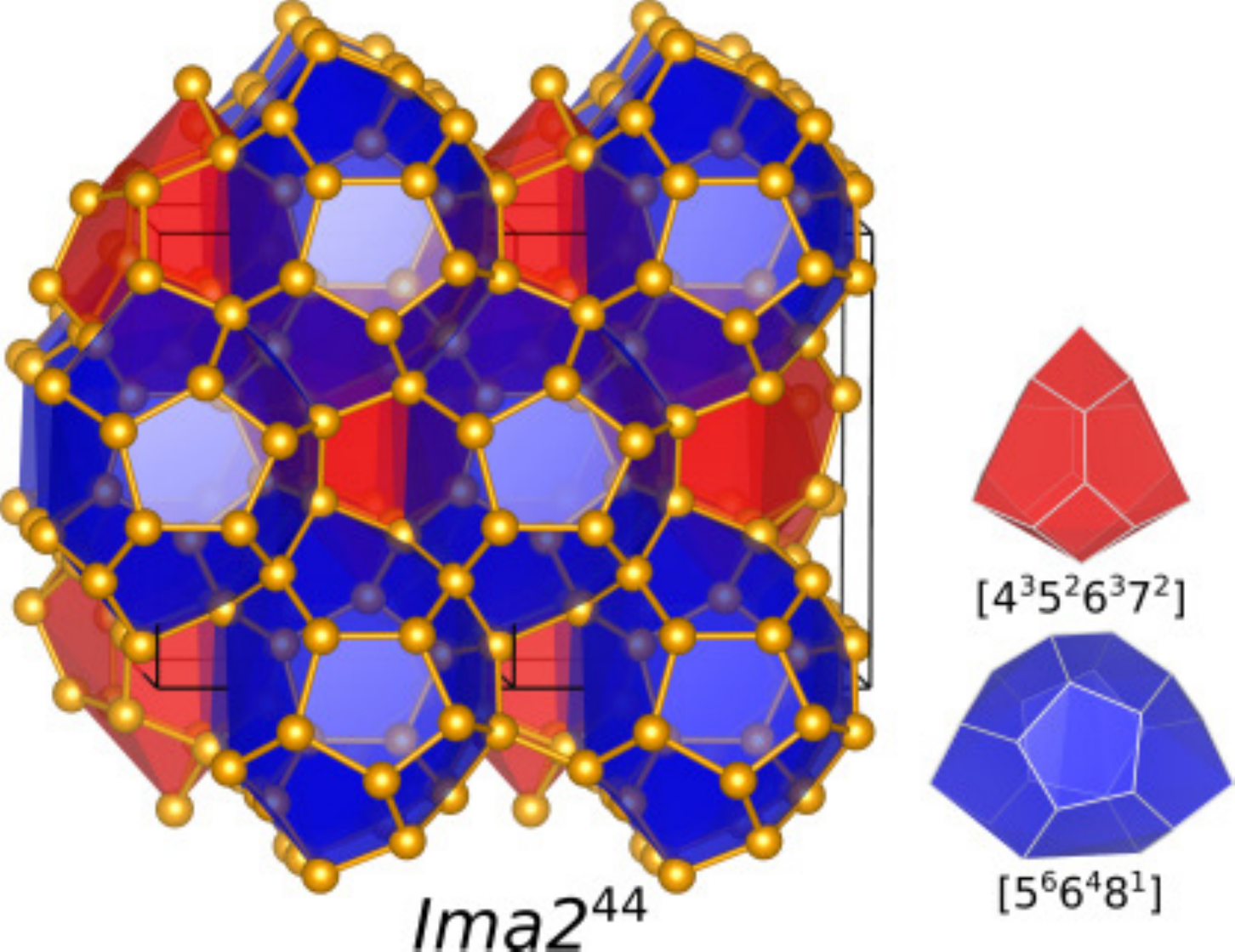}}
\hfill{\includegraphics[height=0.16\textwidth]{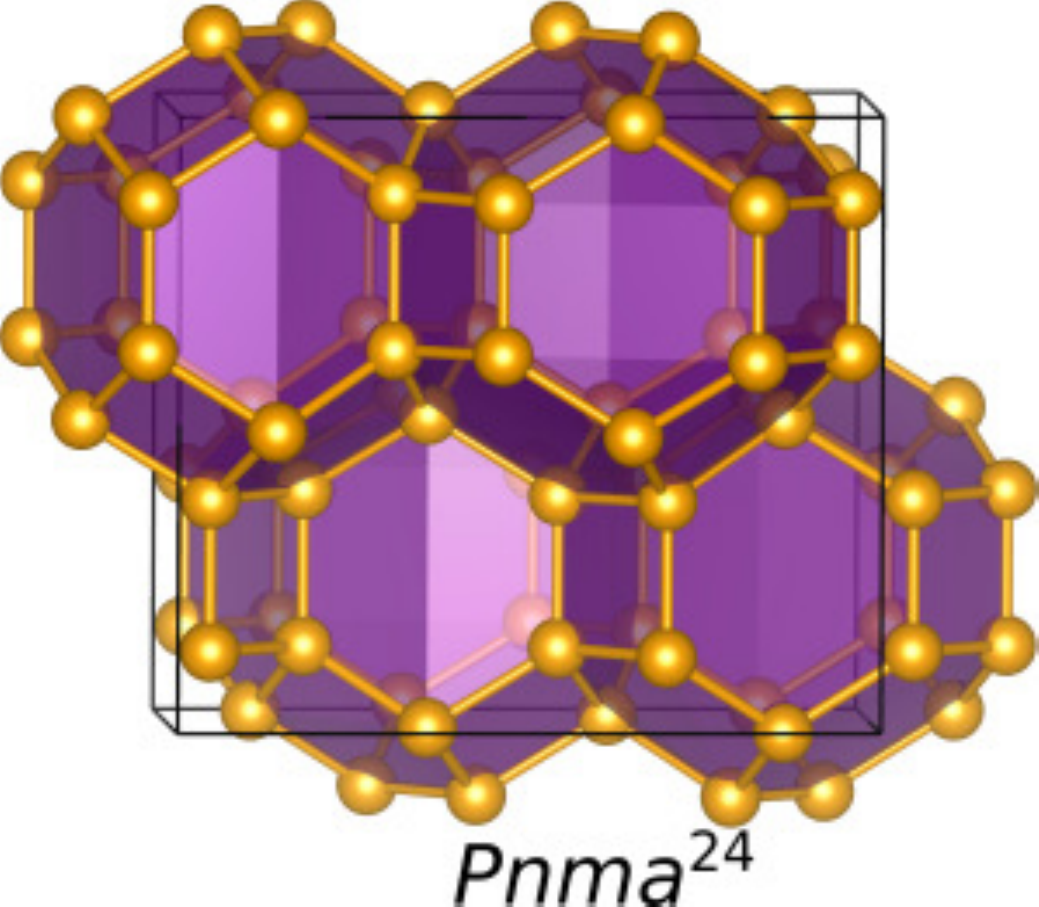}}
\hfill{\includegraphics[height=0.16\textwidth]{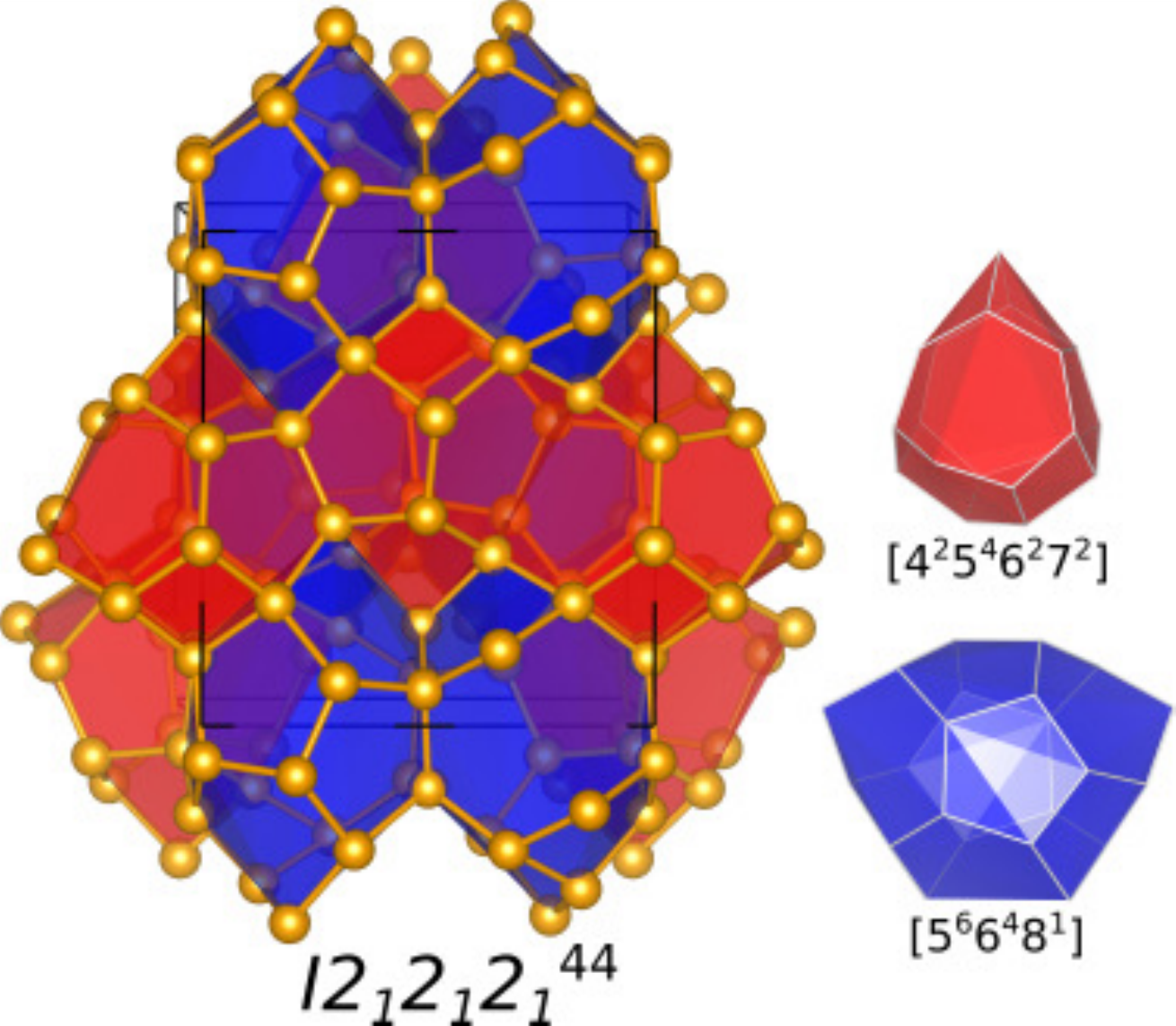}}
\hfill{\includegraphics[height=0.16\textwidth]{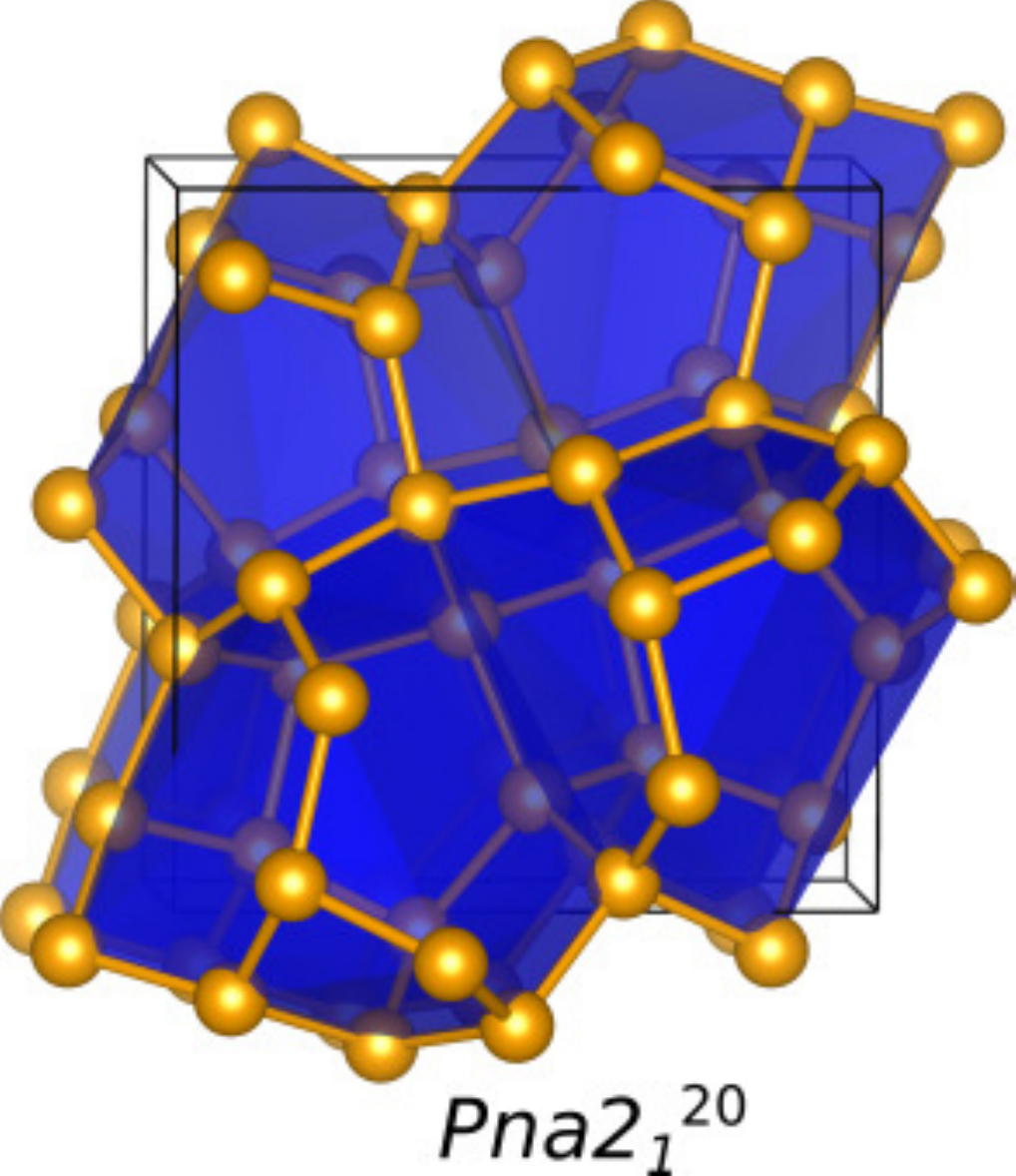}}
\hfill{\includegraphics[height=0.16\textwidth]{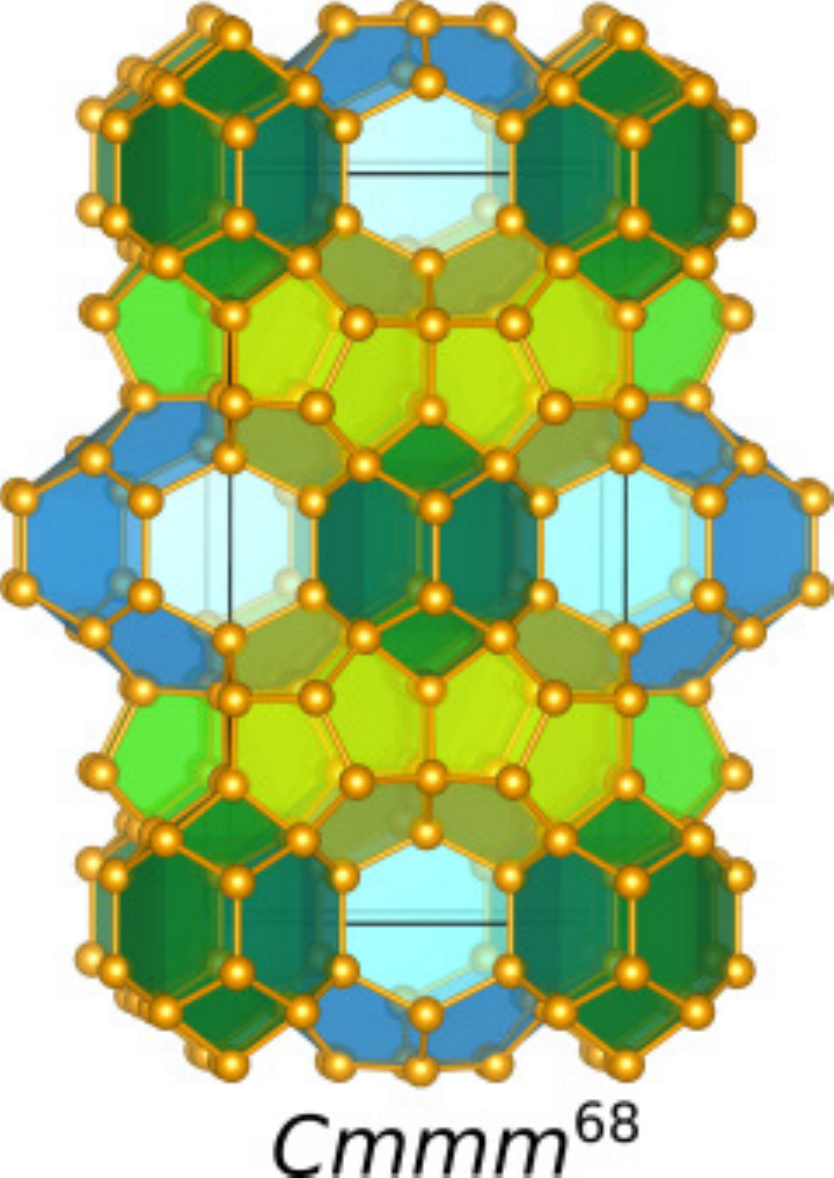}}
\hfill{\includegraphics[height=0.16\textwidth]{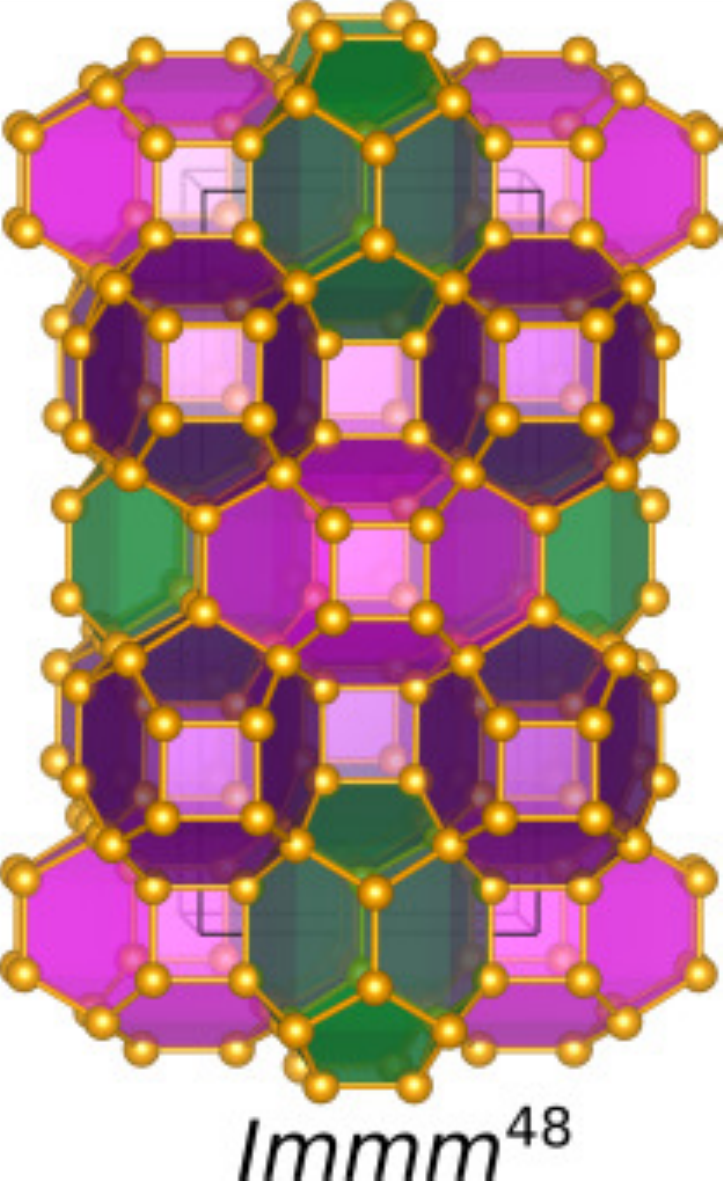}}
\hfill{\includegraphics[height=0.16\textwidth]{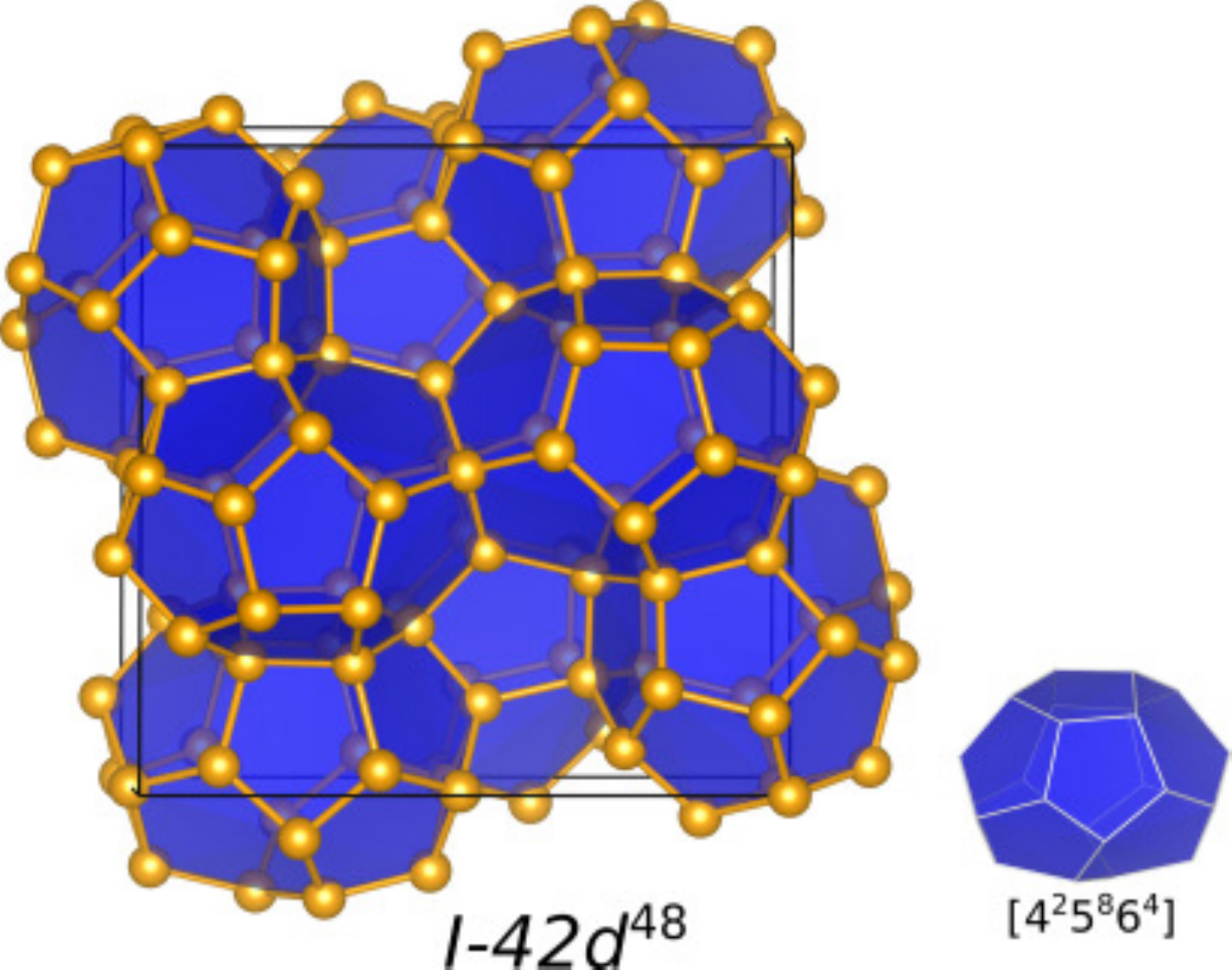}}
\hfill{\includegraphics[height=0.16\textwidth]{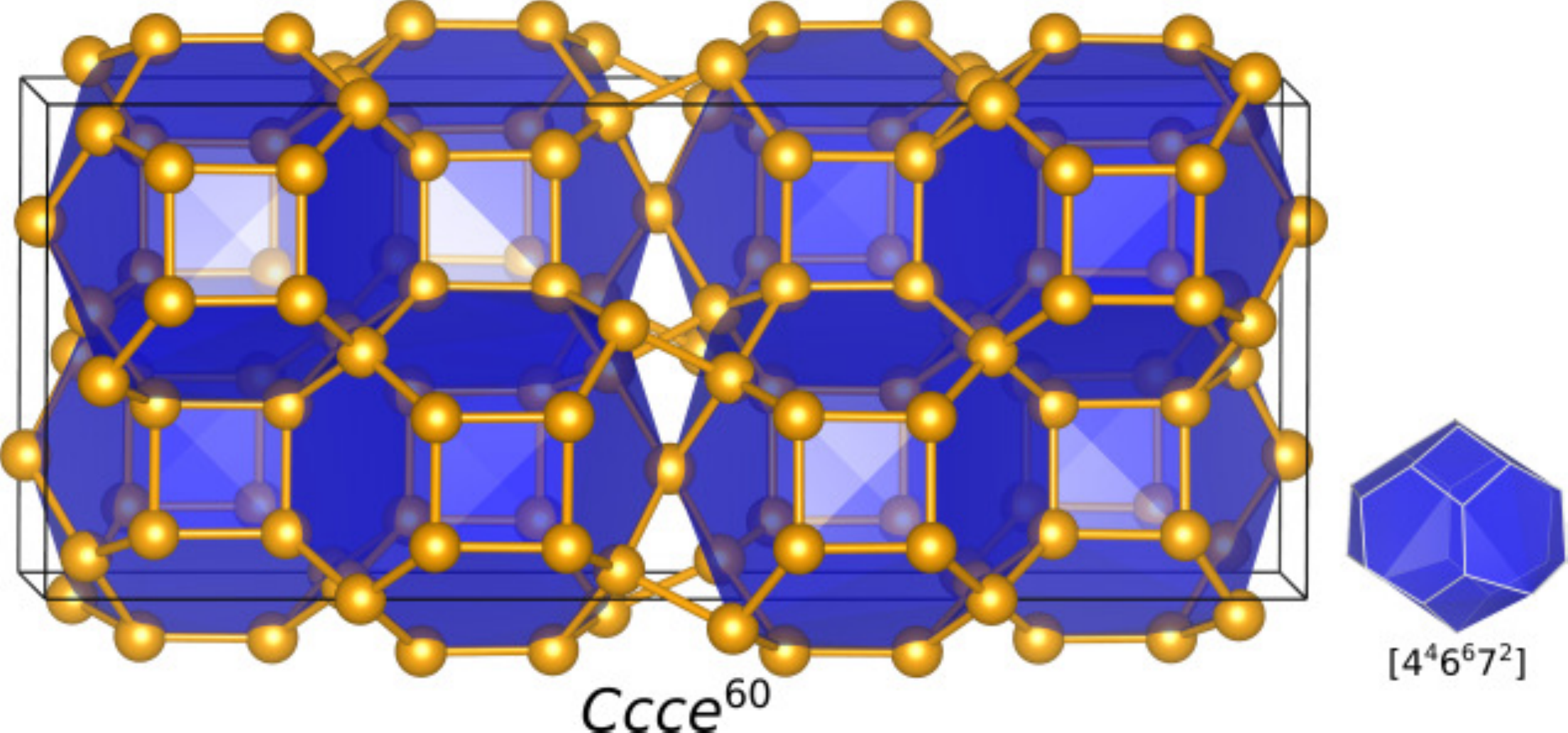}}
\hfill{\includegraphics[height=0.16\textwidth]{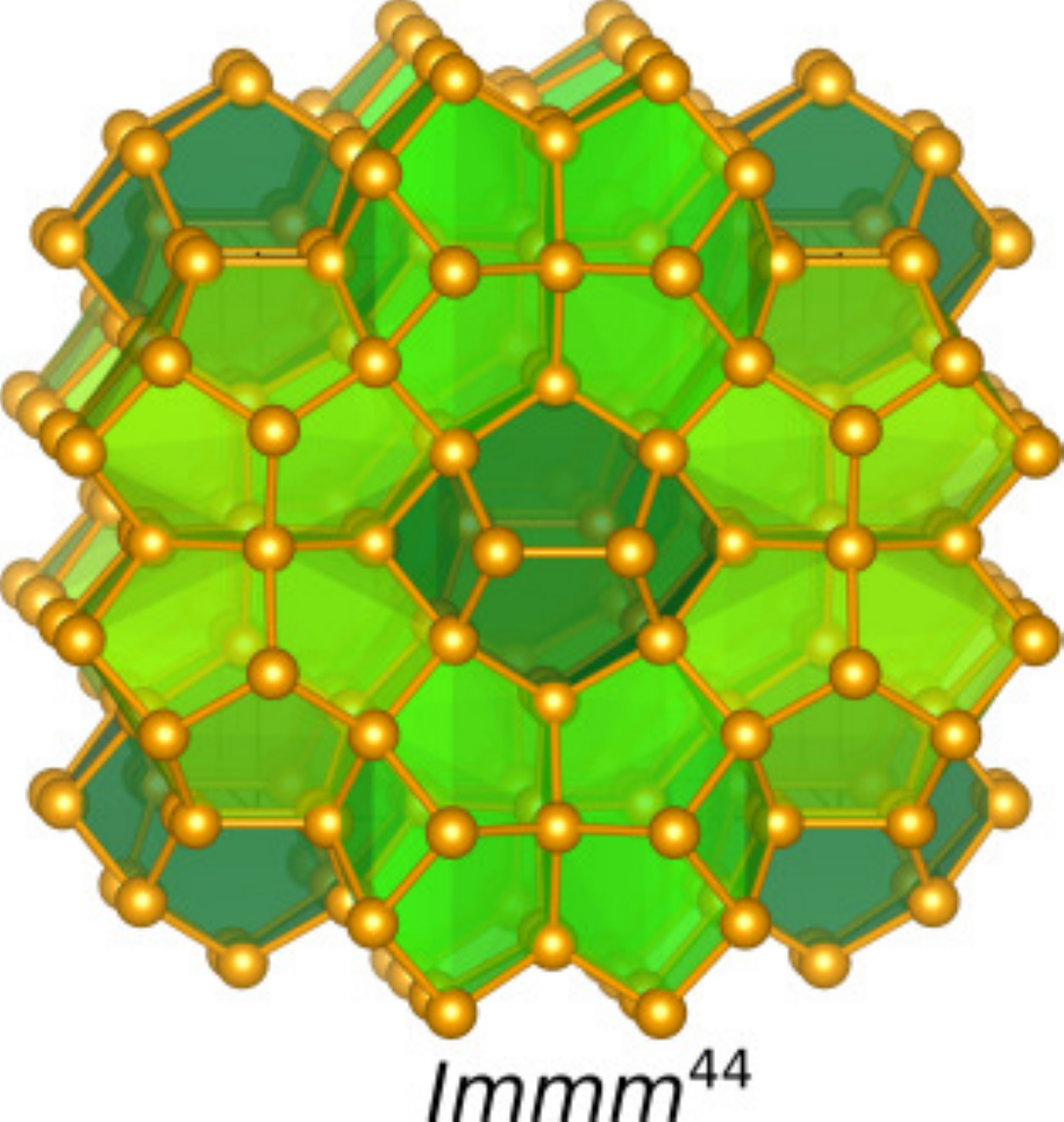}}
\hfill{\includegraphics[height=0.16\textwidth]{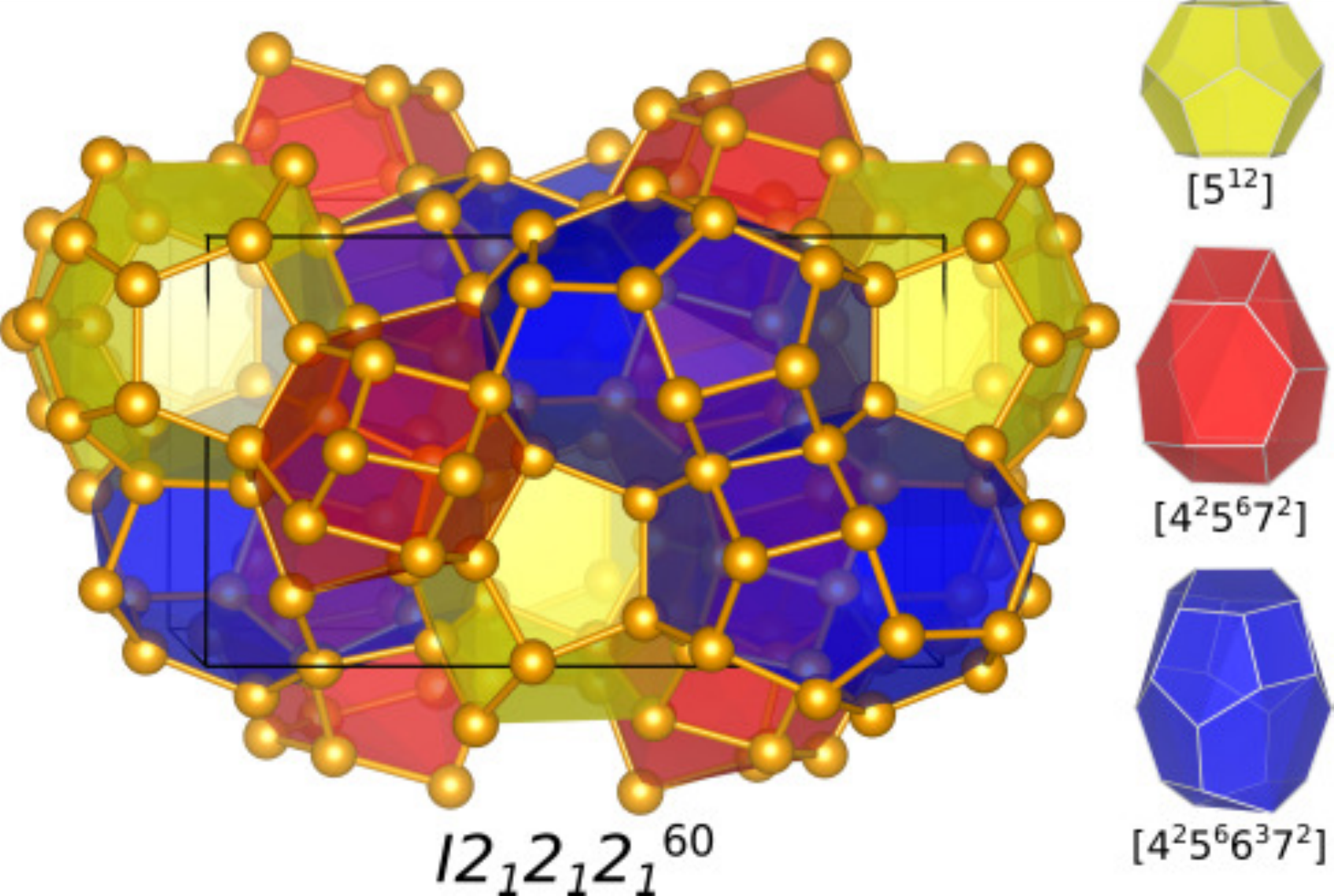}}
\hfill{\includegraphics[height=0.16\textwidth]{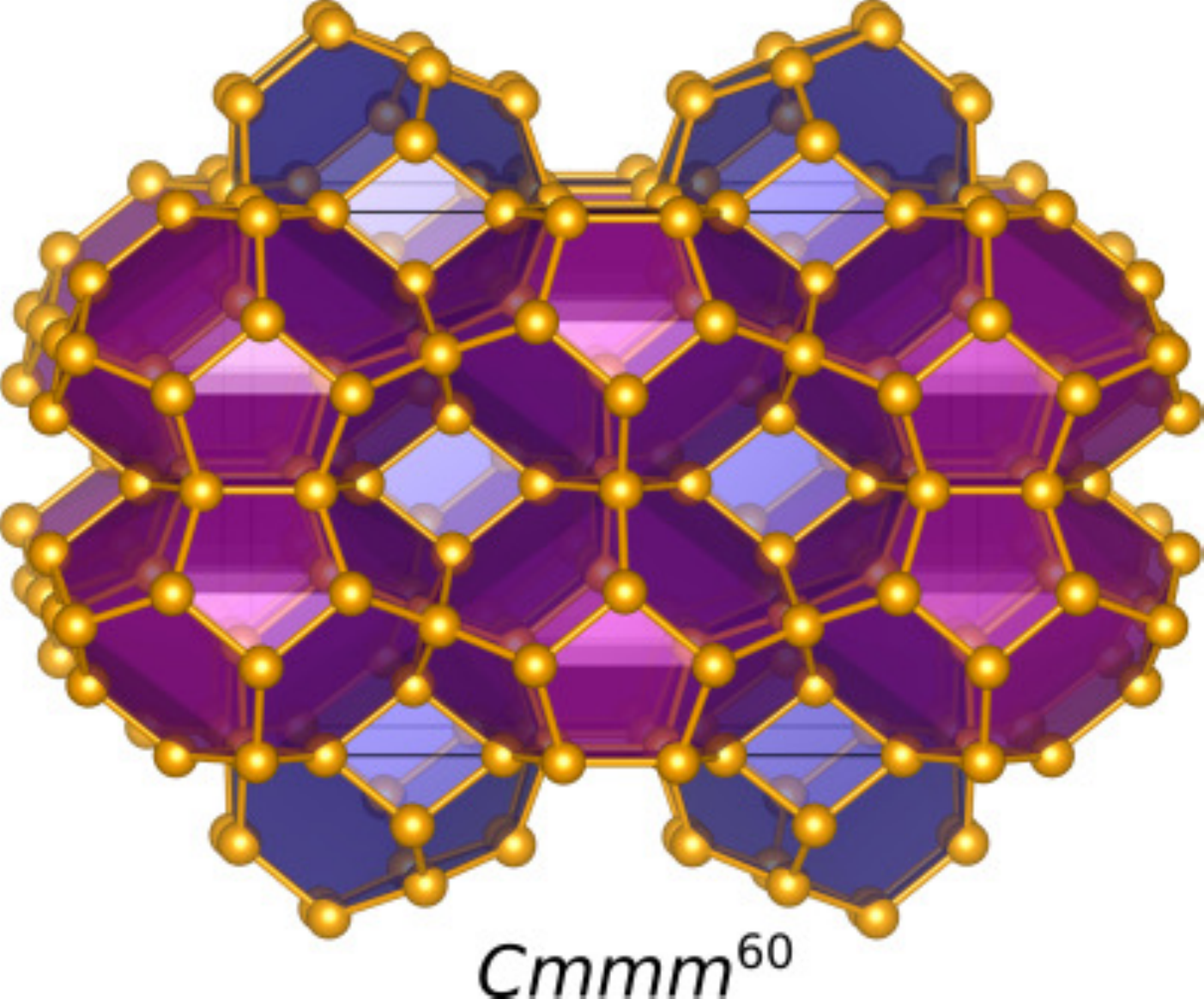}}
\hfill{\includegraphics[height=0.16\textwidth]{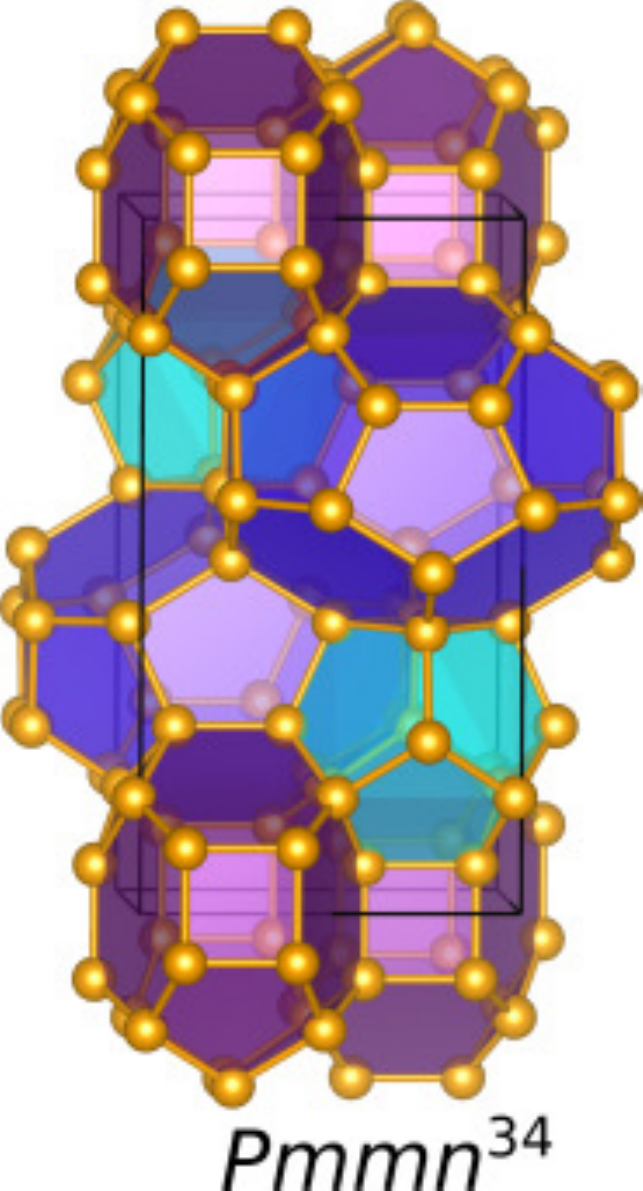}}
\hfill{\includegraphics[height=0.16\textwidth]{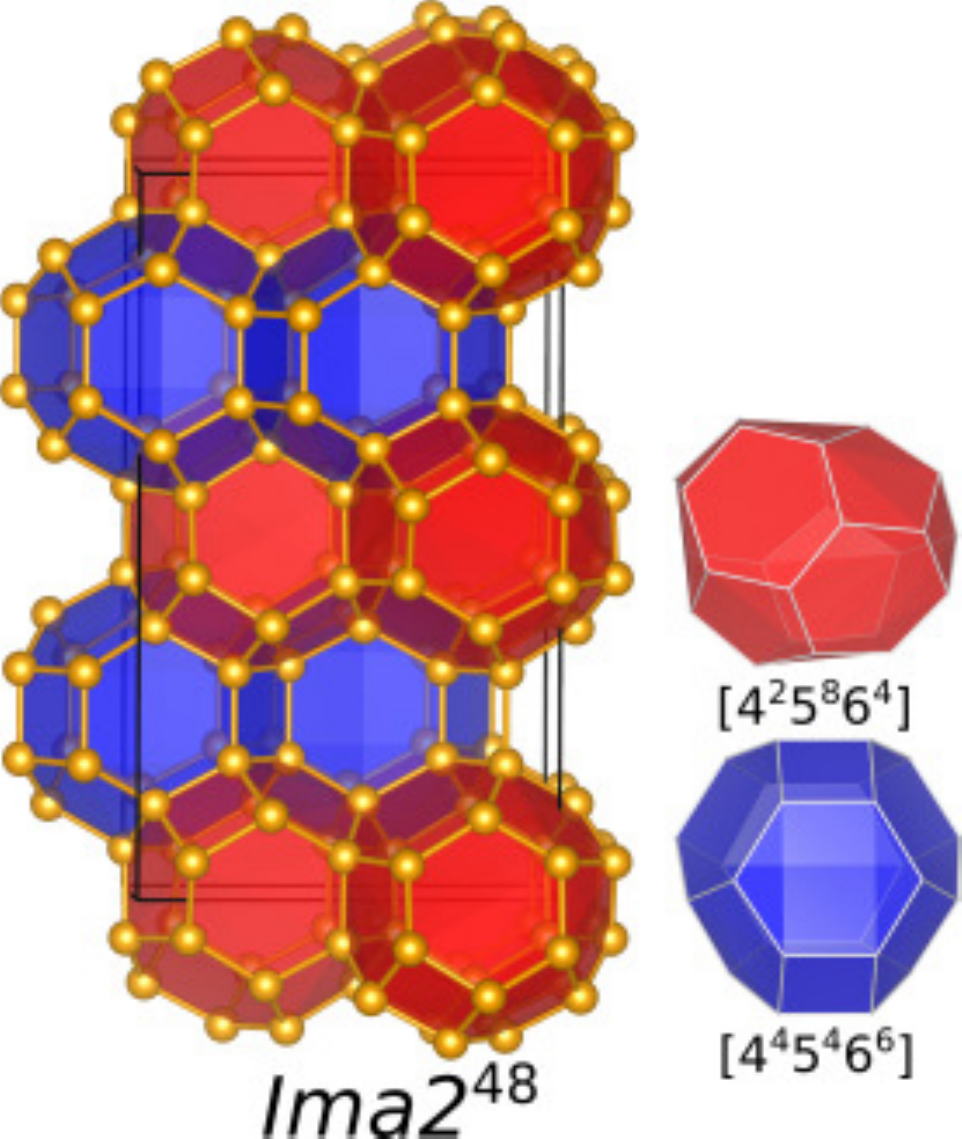}}
\hfill{\includegraphics[height=0.16\textwidth]{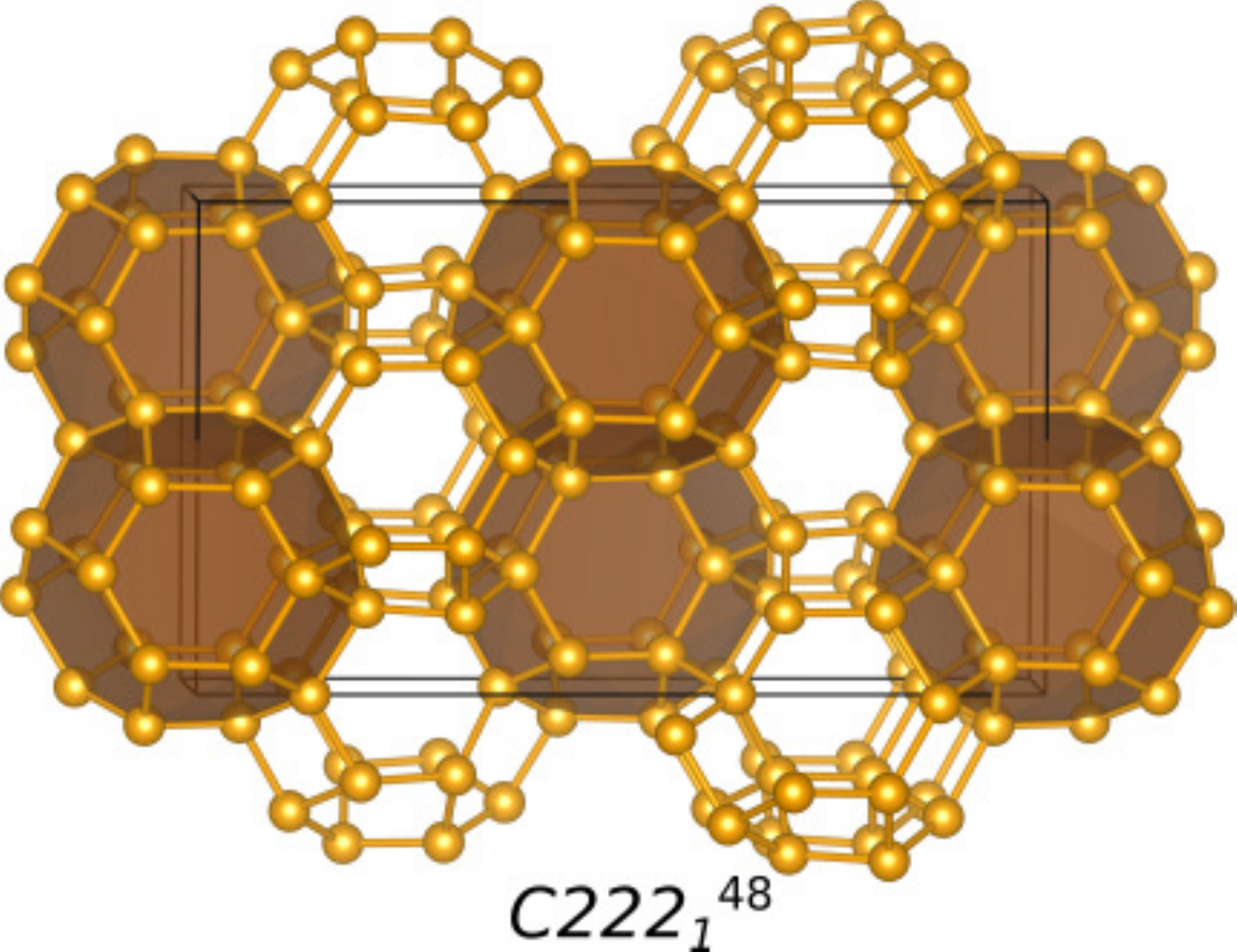}}
\hfill{\includegraphics[height=0.16\textwidth]{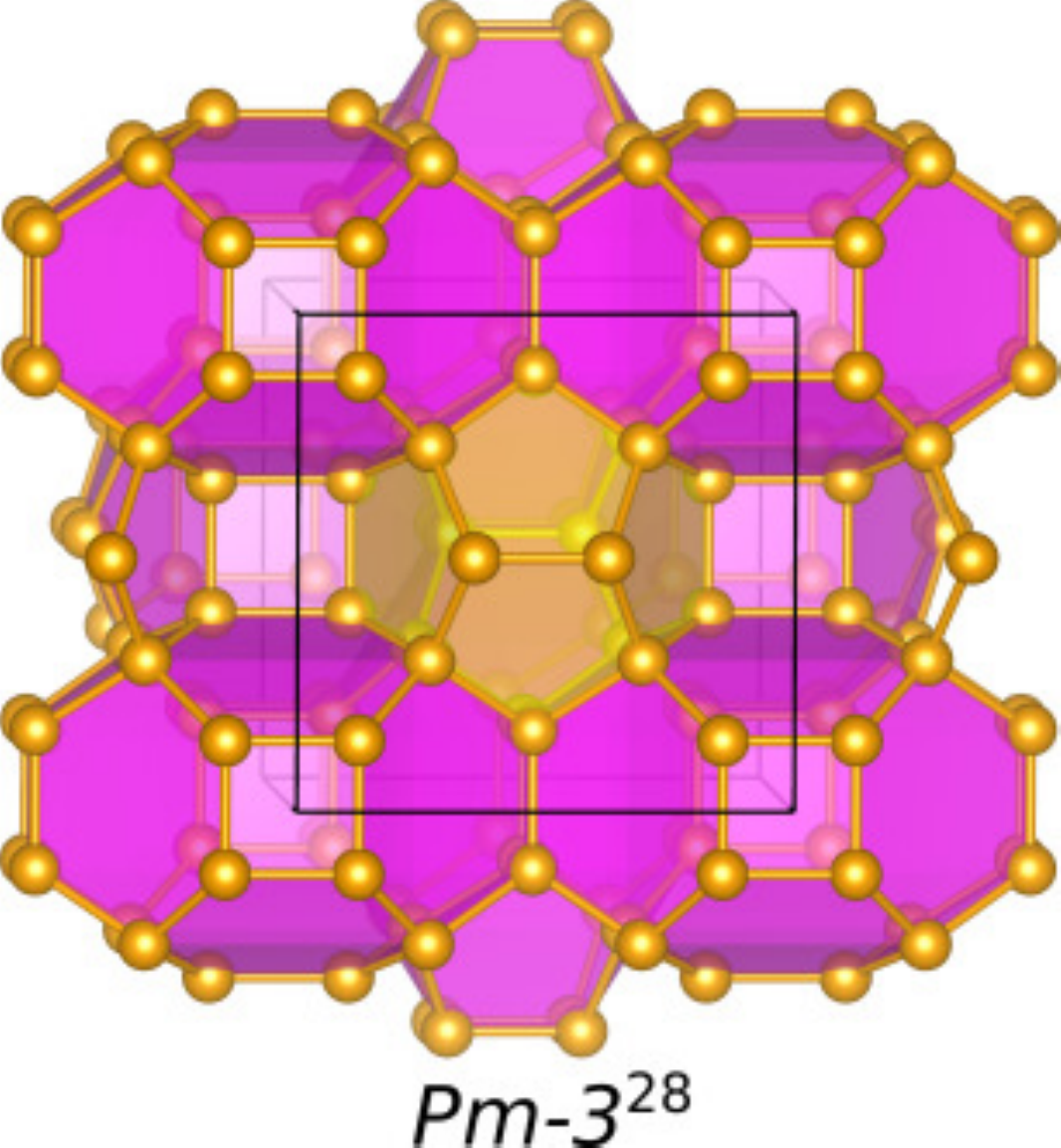}}
\hfill{\includegraphics[height=0.16\textwidth]{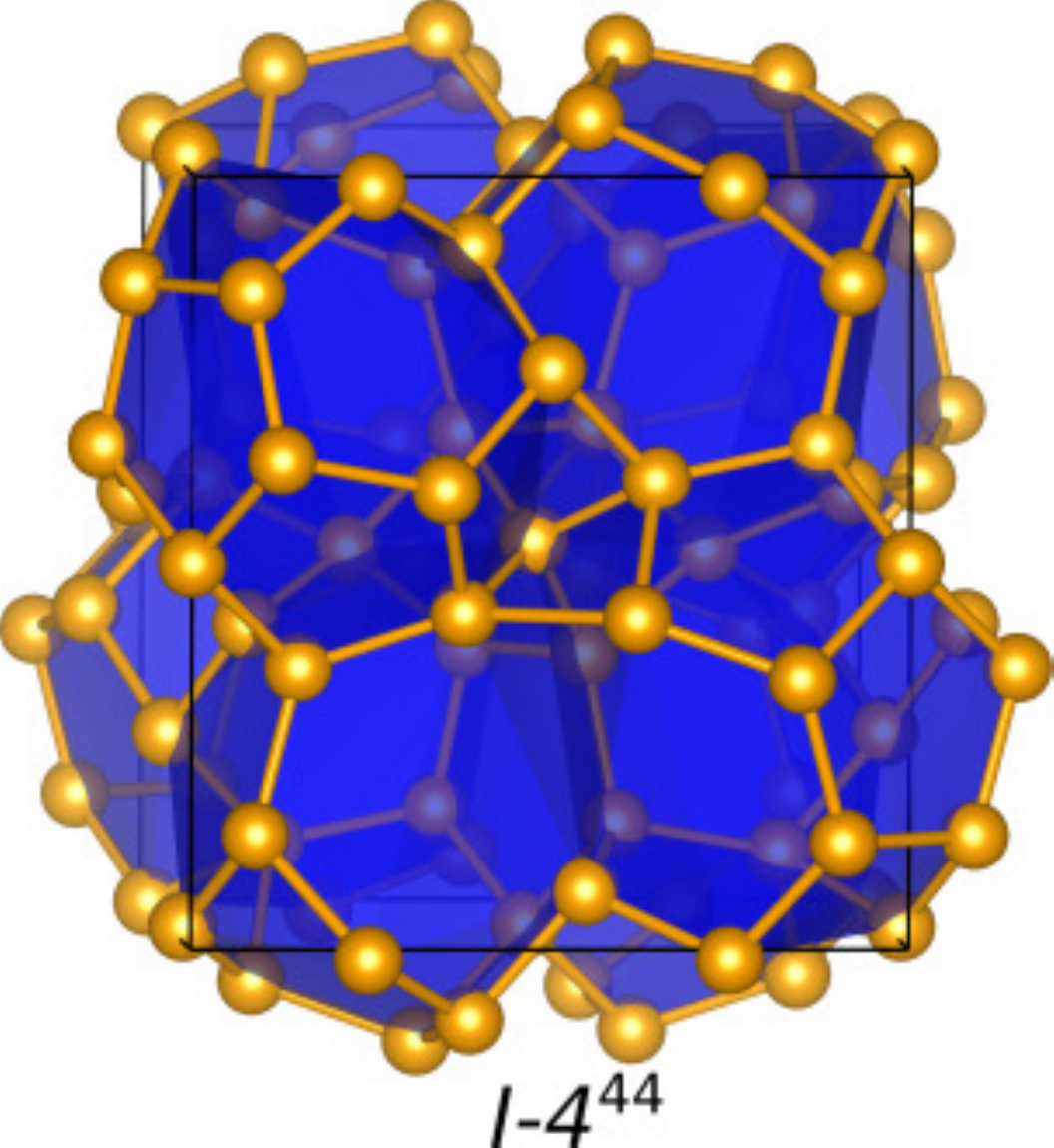}}
\hfill{\includegraphics[height=0.16\textwidth]{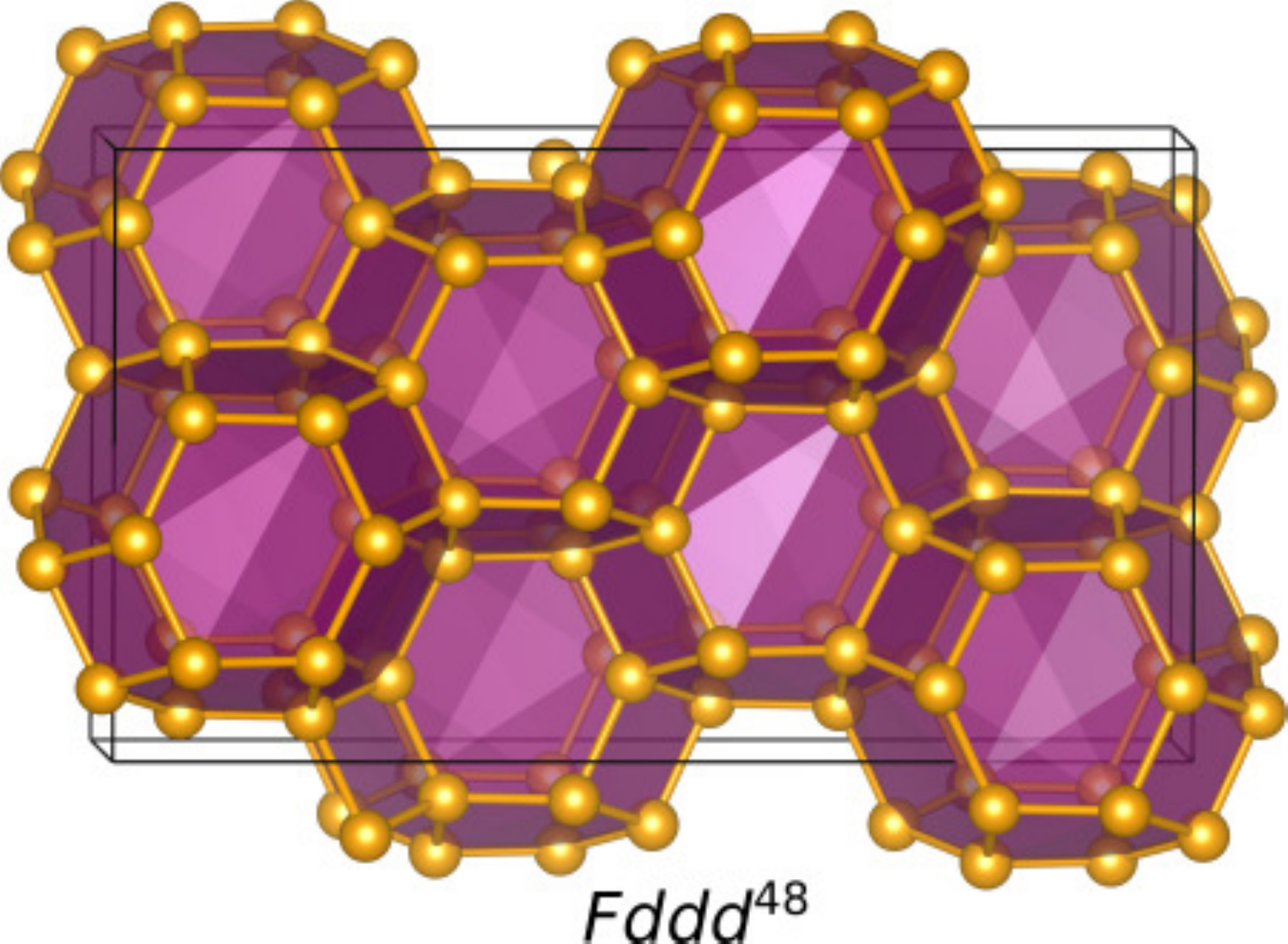}}
\hfill{\includegraphics[height=0.16\textwidth]{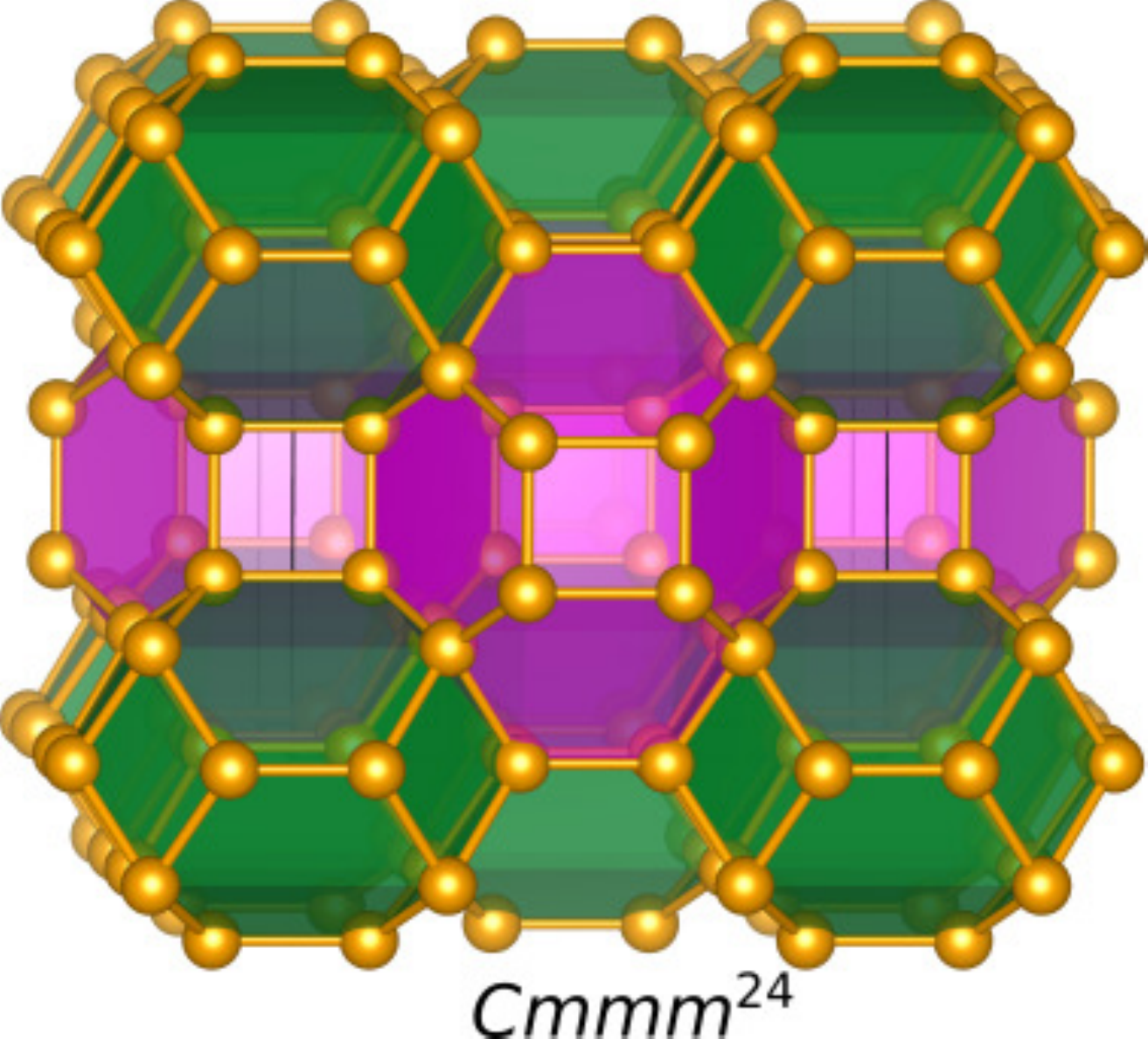}}
\hfill{\includegraphics[height=0.16\textwidth]{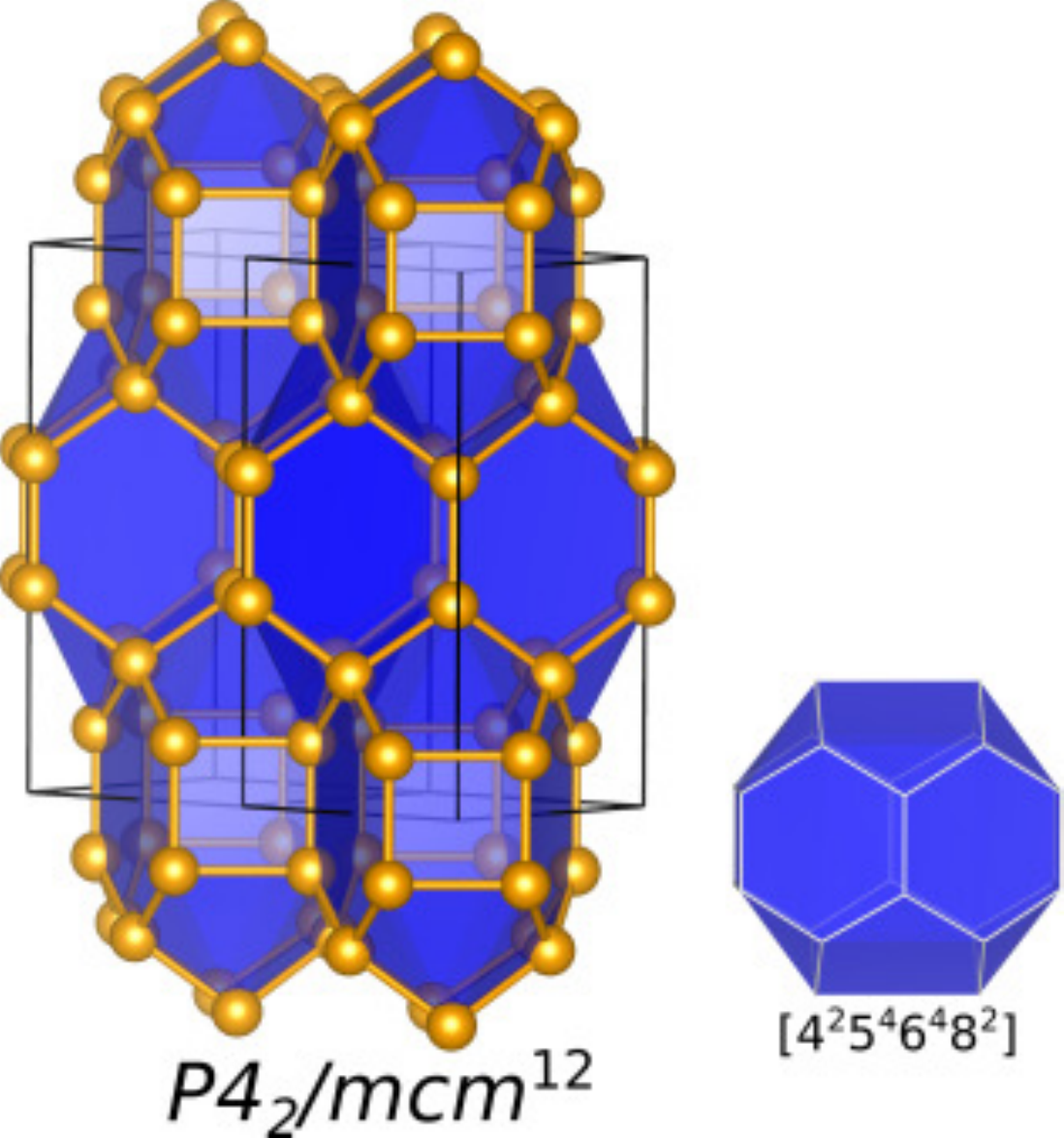}} \hfill
\label{fig:structures}                                                                                                    
 \end{figure*}

Figure~\ref{fig:structures} and Table~\ref{tab:energies} show a selection of 44 low-energy allotropes consisting of cage-like voids which were characterized in detail. Most structures are made from interlinked polyhedral building blocks of 41 different kinds, which can be categorized into polyhedra with planar or nearly planar faces (first two panels in Figure~\ref{fig:polyhedra_0}) or with at least one curved face (three last panels in Figure~\ref{fig:polyhedra_0}). The stability of the individual cages was evaluated in vacuum with the BigDFT wavelet code~\cite{BigDFT2008} together with Hartwigsen-Goedecker-Hutter pseudopotentials~\cite{Pseudopotential}. The calculations were performed imposing zero boundary conditions on the wave-functions and the truncated bonds were passivated with hydrogen atoms. Cages containing 4-rings were found to be energetically unfavorable due to the strong strain on the bond angles, the cohesive energy showing a strong correlation with the number of 4-rings in the cages, indicated by a high correlation coefficient of 0.89. On the other hand, the pentagonal dodecahedra $[5^{12}]$ is the least strained and thus energetically favorable. To some extent, this trend carries on to the solid, where the content of 4-rings increases with the energy of the phases with a correlation coefficient of 0.57. On the other hand, $[5^{12}]$  appears as often as 15 times among 44 phases. Two phases, namely C30~($Pna2_1^{ 20} $) and C42~$I\text{-}4 ^{ 44}$, could not be clearly decomposed into polyhedra, although they have an open framework structure of $sp^3$ bonded silicon atoms.


\begin{table*}[h!]
  \caption{Structural data of the allotropes containing cage-like building blocks. Column 1 contains our structural label, followed by the space group in the Hermann-Mauguin notation with superscripts indicating the number of atoms in the conventional cell. Column 3 contains the characterization of the building blocks, with the different maximum guest molecule radii in the cavities in column 4. Columns 5, 6, and, 7 contain the energy per atom with respect to $d$-Si, the volume per atom and the bulk modulus, respectively. The last column contains the corrected PBE band-gap. ``D'' indicates a direct or quasi-direct band-gap.}
\tiny
\label{tab:energies}
\begin{tabular} {l l l l l l l l}
Label & Space-group & Building blocks & Guest radii (\AA) & Energy (eV) & Vol (\AA$^3$)& B$_0$ & Gap (eV)\\
 \hline \\
C01  & $R\text{-}3m^{204} $ & $ [4^{3}5^{6}6^{3}]_6[5^{12}]_{14}[5^{12}6^{4}]_6[5^{12}6^{8}]_3                                                  $& 2.0,2.1,2.8,3.1         &  0.0687 &   23.82 & 75.54& 1.87 D \\ 
C02 & $Fmmm        ^{ 48} $ & $ [4^{2}5^{8}6^{4}]_4                                                                                             $& 1.7                     &  0.0835 &   22.20 & 80.09& 1.39  \\ 
C03  & $C2/m       ^{ 80} $ & $ [4^{2}5^{8}6^{1}]_4[5^{12}]_2[5^{12}6^{2}]_4[5^{12}6^{4}]_4                                                     $& 1.9,2.1,2.2,2.6         &  0.0901 &   23.62 & 75.62& 1.58 D \\ 
C04  & $Amm2       ^{212} $ & $ [4^{2}5^{8}6^{1}]_4[4^{4}5^{8}6^{3}]_2[5^{12}]_{16}[5^{12}6^{2}]_4[5^{10}6^{2}]_2[5^{12}6^{3}]_2[5^{12}6^{4}]_8 $& 1.8,2.1,2.3,2.5,2.7,2.8 &  0.1008 &   23.45 & 75.89& 0.67  \\ 
C05  & $F222       ^{128} $ & $ [5^{4}6^{2}7^{4}]_6[5^{4}7^{4}]_4[6^{2}7^{2}8^{2}]_6                                                            $& 1.7,2.0,2.2             &  0.1062 &   21.65 & 81.51& 1.60  \\ 
C06  & $R\text{-}3m^{128} $ & $ [4^{3}5^{6}6^{4}]_4[5^{12}]_{14}[5^{12}6^{2}]_6[5^{12}6^{6}]_3                                                  $& 2.0,2.3,2.4,2.9         &  0.1101 &   23.79 & 74.54& 1.74 D \\ 
C07  & $Immm       ^{ 56} $ & $ [4^{2}5^{8}]_4[5^{12}]_2[5^{12}6^{5}]_2                                                                         $& 1.8,2.7,2.8             &  0.1243 &   24.03 & 73.20& 1.55 D \\ 
C08  & $Pmma       ^{ 46} $ & $ [4^{1}5^{10}6^{7}]_2[4^{2}5^{8}6^{2}]_2[4^{3}5^{6}6^{3}]_2[5^{12}]_2                                            $& 1.8,2.0,2.9             &  0.1253 &   24.11 & 73.04& 1.48 D \\ 
C09  & $R\text{-}3m^{ 84} $ & $ [4^{3}5^{6}6^{4}_{\mathrm{I}}]_8[5^{12}]_3[5^{8}6^{4}]_6                                                        $& 2.1                     &  0.1265 &   22.55 & 78.05& 1.53  \\ 
C10 & $Cmcm       ^{ 68} $ & $ [4^{2}5^{8}6^{1}]_4[5^{12}]_4[5^{12}6^{5}]_4                                                                     $& 1.7,1.9,2.7             &  0.1271 &   23.85 & 73.92& 1.01  \\ 
C11 & $Ama2       ^{ 68} $ & $ [5^{3}6^{3}7^{3}]_4[5^{6}6^{1}7^{2}]_4[5^{8}8^{2}]_4                                                             $& 1.6,1.8,2.1             &  0.1280 &   22.11 & 79.65& 1.35  \\ 
C12 & $I4_1md     ^{ 48} $ & $ [4^{1}5^{10}6^{2}]_4[4^{3}5^{6}6^{6}]_4                                                                          $& 2.0,2.2                 &  0.1320 &   23.47 & 74.59& 1.77 D \\ 
C13 & $R3m        ^{138} $ & $ [4^{3}5^{6}6^{1}]_3[4^{3}5^{6}6^{4}]_4[4^{6}5^{3}6^{1}7^{3}]_2[5^{12}]_9[5^{12}6^{2}]_3[5^{15}6^{4}7^{3}]_3      $& 1.7,2.0,2.2,2.3,2.9     &  0.1323 &   24.42 & 71.85& 1.62  \\ 
C14 & $F222       ^{128} $ & $ [5^{4}6^{2}7^{4}]_4[5^{4}7^{4}]_8[6^{2}7^{4}8^{2}]_4                                                             $& 1.7,2.0,2.1             &  0.1335 &   21.50 & 81.77& 1.49  \\ 
C15 & $Pmn2_1     ^{ 30} $ & $ [4^{1}5^{6}6^{1}8^{2}]_2[4^{1}6^{4}8^{3}]_2[5^{7}7^{1}8^{1}]_2                                                   $& 1.6,1.9,2.0             &  0.1358 &   22.38 & 78.27& 1.98  \\ 
C16 & $Pmmm       ^{ 40} $ & $ [4^{3}5^{6}6^{3}]_2[5^{12}]_1[5^{12}6^{2}]_2[5^{12}6^{8}]_1                                                      $& 2.1,2.2,2.9             &  0.1437 &   24.05 & 73.09& 0.83  \\ 
C17 & $Pbcn       ^{ 16} $ & $ [4^{2}5^{4}7^{4}]_4                                                                                              $& 1.6                     &  0.1469 &   22.62 & 76.85& 0.94  \\ 
C18 & $Pmn2_1     ^{ 30} $ & $ [3^{2}4^{1}5^{2}6^{6}8^{1}]_2[4^{1}5^{6}6^{3}9^{1}]_2[5^{7}7^{1}8^{1}]_2                                         $& 1.5,1.6,1.7             &  0.1491 &   22.25 & 78.73& 1.56  \\ 
C19 & $Pmmm       ^{ 24} $ & $ [4^{2}5^{8}6^{4}]_2[4^{4}5^{4}6^{8}]_1[5^{12}]_1                                                                 $& 1.7,1.8,2.2             &  0.1491 &   23.49 & 74.21& 1.21  \\ 
C20 & $Cccm       ^{164} $ & $ [4^{1}5^{10}6^{2}]_8[4^{2}5^{8}6^{5}]_8[4^{3}5^{6}6^{4}]_8[5^{12}6^{2}]_4                                        $& 1.9,2.1,2.2,2.3         &  0.1498 &   23.59 & 74.03& 0.68  \\ 
C21 & $Amm2       ^{ 40} $ & $ [4^{2}5^{8}6^{4}]_2[4^{3}5^{4}6^{4}8^{1}]_4 [4^{4}6^{4}8^{2}]_2                                                  $& 1.7,1.8,1.9             &  0.1504 &   23.59 & 73.89& 1.68  \\ 
 C22 & $Ima2       ^{ 32} $ & $ [4^{2}5^{4}6^{1}8^{2}]_4[4^{2}5^{4}7^{4}]_6                                                                      $& 1.6,1.7                 &  0.1543 &   22.64 & 76.55& 0.97  \\ 
 C23 & $R\text{-}3m^{ 72} $ & $ [4^{2}5^{8}6^{4}_{\mathrm{III}}]_9[4^{6}6^{8}]_3                                                                 $& 1.8,2.4                 &  0.1587 &   23.80 & 72.83& 1.33  \\ 
 C24 & $I4/m       ^{ 60} $ & $ [4^{2}5^{8}6^{4}_{\mathrm{I}}]_8[4^{2}5^{8}6^{4}_{\mathrm{II}}]_2                                                $& 1.9,2.1                 &  0.1621 &   23.46 & 74.10& 1.65  \\ 
 C25 & $Pnma       ^{ 16} $ & $ [4^{2}5^{4}6^{2}8^{2}]_4                                                                                         $& 1.7                     &  0.1637 &   22.68 & 76.37& 0.95  \\ 
 C26 & $Ima2       ^{ 44} $ & $ [4^{3}5^{2}6^{3}7^{2}]_4[5^{6}6^{4}8^{1}]_4                                                                      $& 1.9,2.0                 &  0.1646 &   22.69 & 76.37& 1.26  \\ 
 C27 & $Pnma       ^{ 24} $ & $ [4^{4}5^{4}6^{6}]_4                                                                                              $& 2.1                     &  0.1670 &   24.33 & 71.13& 1.69 D \\ 
 C28 & $I2_12_12_1 ^{ 44} $ & $ [4^{2}5^{4}6^{2}7^{2}]_4[5^{6}6^{4}8^{1}]_4                                                                      $& 1.8,2.0                 &  0.1677 &   22.44 & 77.63& 1.38  \\ 
 C29 & $Pna2_1     ^{ 20} $ & $                                                                                                                  $& 1.6                     &  0.1705 &   22.61 & 76.78& 1.78  \\ 
 C30 & $Cmmm       ^{ 68} $ & $ [4^{2}5^{8}6^{12}]_2[4^{4}5^{4}6^{3}]_4[4^{4}5^{4}6^{4}]_2[5^{12}]_4                                             $& 1.8,1.9,2.0,3.3         &  0.1708 &   24.80 & 69.85& 1.04  \\ 
 C31 & $Immm       ^{ 48} $ & $ [4^{4}5^{4}6^{4}]_2[4^{4}5^{4}6^{6}]_4[4^{4}5^{4}6^{8}]_2                                                        $& 1.6,2.0,2.3             &  0.1717 &   24.27 & 71.52& 1.18 D \\ 
 C32& $I\text{-}42d^{ 48} $ & $ [4^{2}5^{8}6^{4}]_8                                                                                             $& 1.6                     &  0.1773 &   22.98 & 75.12& 0.99 D \\ 
 C33& $Ccce       ^{ 60} $ & $ [4^{2}6^{6}7^{2}]_8                                                                                              $& 2.2                     &  0.1781 &   23.48 & 73.41& 0.73  \\ 
 C34& $Immm       ^{ 44} $ & $ [4^{3}5^{6}6^{3}]_4[4^{4}5^{4}6^{3}]_4[4^{4}5^{4}6^{4}]_2                                                        $& 1.6,2.0,2.1             &  0.1800 &   23.04 & 75.26& 1.08 D \\ 
 C35& $I2_12_12_1 ^{ 60} $ & $ [4^{2}5^{6}6^{3}7^{2}]_4[4^{2}5^{6}7^{2}]_4[5^{12}]_4                                                            $& 1.7,2.0                 &  0.1834 &   22.81 & 75.77& 1.50 D \\ 
 C36& $Cmmm       ^{ 60} $ & $ [4^{3}5^{6}6^{5}]_4[4^{4}5^{4}6^{8}_{\mathrm{II}}]_4                                                             $& 2.1,2.5                 &  0.1899 &   24.05 & 71.95& 0.62  \\ 
 C37& $Pmmn       ^{ 34} $ & $ [4^{1}5^{10}6^{6}]_2[4^{3}5^{6}]_2[4^{4}5^{4}6^{6}]_2                                                            $& 1.3,1.8,2.7             &  0.1954 &   24.48 & 70.52& 1.20  \\ 
 C38& $Ima2       ^{ 48} $ & $ [4^{4}5^{4}6^{6}]_4[4^{2}5^{8}6^{4}]_4                                                                           $& 1.8,2.1                 &  0.1960 &   23.77 & 72.17& 1.20  \\ 
 C39& $C222_1     ^{ 48} $ & $ [4^{2}5^{6}6^{4}]_4                                                                                              $& 1.7                     &  0.1971 &   22.85 & 75.48& 1.30  \\ 
 C40& $Pm\text{-}3^{ 28} $ & $ [4^{4}5^{4}6^{8}]_3[5^{12}]_1                                                                                    $& 2.0,2.3                 &  0.2010 &   24.48 & 70.45 & 0.89  \\ 
 C41& $I\text{-}4 ^{ 44} $ & $                                                                                                                  $& 1.9                     &  0.2012 &   22.78 & 75.63& 1.39  \\ 
 C42& $Fddd       ^{ 48} $ & $ [4^{3}5^{4}6^{6}]_8                                                                                              $& 2.3                     &  0.2317 &   23.99 & 70.53& --  \\ 
 C43& $Cmmm       ^{ 24} $ & $ [4^{4}5^{4}6^{8}]_2[4^{4}5^{4}6^{4}]_2                                                                           $& 1.7,2.5                 &  0.2542 &   24.20 & 69.74& 0.75  \\ 
 C44& $P4_2/mcm   ^{ 12} $ & $ [4^{2}5^{4}6^{4}8^{2}]_2                                                                                         $& 1.6                     &  0.2915 &   22.84 & 72.50& 0.68  \\ 
\end{tabular}
\end{table*}

\begin{figure*}[h!]
\caption{Energies as a function of volume/atom of the structures C01-C44 with respect to $d$-Si, together with the silicon clathrates type I and type II, and the predicted silicon allotrope Si20T which exhibits the best absorption to date~\cite{xiang_towards_2013}.}            
\centering
\setlength{\unitlength}{1cm}
{\includegraphics[width=0.75\textwidth]{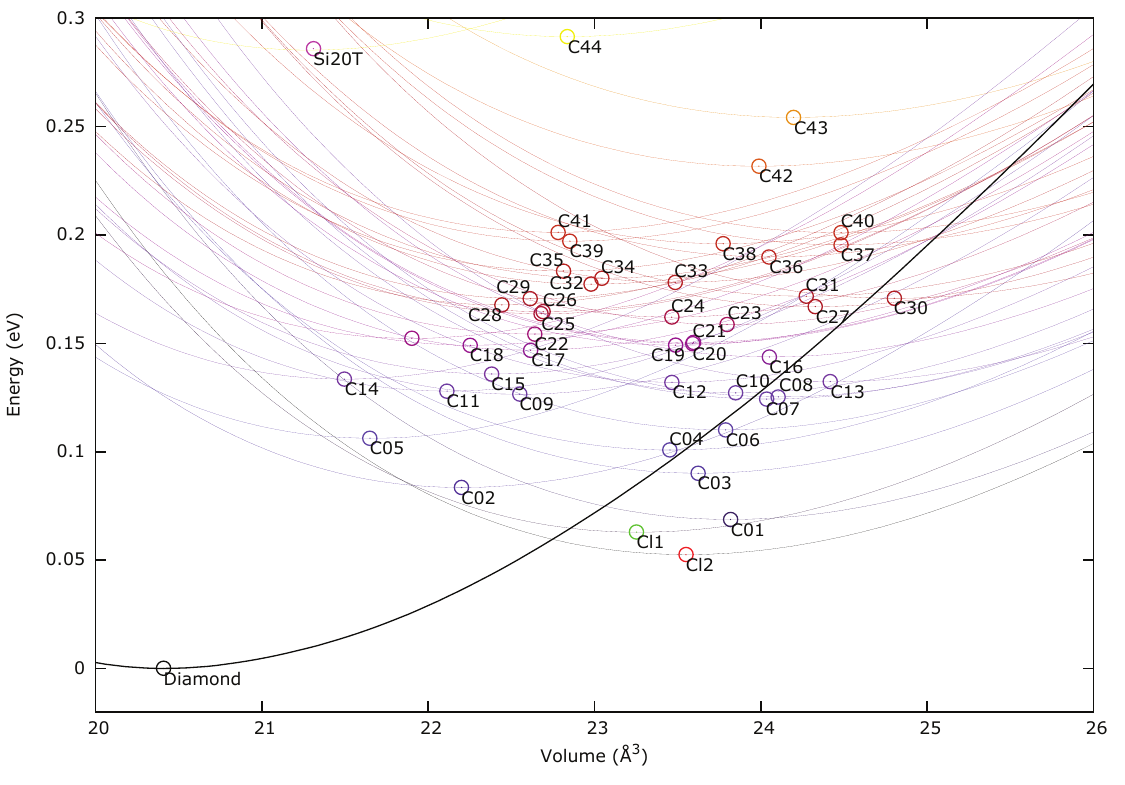}}
\label{fig:energy_vol}
 \end{figure*}

Phonon calculations were carried out to assess the dynamical lattice stability for every phase. No imaginary phonon modes were found, thus confirming their structural stability (see supplemental materials). From the  energetic point of view, all structures are less favorable than $d$-Si or the most stable silicon clathrate of type II and are thus metastable. The relative energies with respect to $d$-Si are plotted as a function of the atomic volume for all 44 phases in Figure~\ref{fig:energy_vol}. However, at least the 20 lowest energy structures ranging from C01 to C20 are below 150~meV/atom in energy relative to $d$-Si and are thus likely to be thermodynamically accessible in experiments. Furthermore, 8 of those structures have energies below or very close to the $d$-Si line at their specific equilibrium volumes, namely C01, C03, C04, C06, C07, C08, C10 and C13, thus representing the most promising set of candidates for future experimental synthesis. Furthermore, the bulk moduli $B_0$ were computed through a fit to the Murnaghan equation of states (see table~\ref{tab:energies}) and are consistently lower than the value for $d$-Si of 87.9~GPa.

Clathrate compounds are characterized by a framework of an open host structure with cavities occupied by guest molecules or atoms, and naturally the size of the cavities determines the maximum size of the guest atoms hosted by the crystal lattice. Although Na is the most prominent guest atom in silicon clathrates, clathrate materials can host various guest atoms which interact weakly with the host atoms based on the Zintl concept and profoundly change the electronic properties~\cite{yamanaka_high-pressure_2000,rowe_thermoelectrics_2005,yamanaka_high-pressure_2000}. To asses the size of guest molecules that can be hosted by the cavities, we estimated the radii of the largest possible spheres placed at the center of each polyhedra without distorting the host structure. All silicon atoms were assumed to be rigid spheres with a covalent radius of 1.11\AA. Since the average bond-length in the solid is slightly larger than twice the covalent radius ($\approx$2.38~\AA) and all atoms are fully coordinated with no bonds pointing towards the center of the voids, we can safely assume that this value is an upper limit. The smallest cavity of type $[4^35^6]$ can thus host guests with a radius of merely 1.3~\AA, whereas the largest could enclose molecules with radii of 3.3~\AA~corresponding to the $[4^25^86^{12}]$ cage. The average guest radius for all structures is listed in table~\ref{tab:energies} is 2.0~\AA. Nevertheless, the estimated spherical radii for the guest atoms should be considered with care and should only be taken as a guide for the maximal size of inert guest molecules.

Most structures are semi-conducting with gaps in the range between about 0.6 and 2.0~eV as shown in table~\ref{tab:energies}. Several structures exhibit a direct or quasi-direct band gap, namely C01, C03, C06, C07, C08, C12, C28, C32, C33, C35 and C36 (see table~\ref{tab:energies} and I03, I11, I12, I24, I42, I46 in the supplemental material). The most promising candidates for photovoltaic applications are thus C03, C07, C08, C32 and C36, which have gaps close to the optimal value for the Shockley-Queisser limit of $\approx 1.4$~eV.

\begin{figure*}[h!]            
\caption{Absorption spectra of selected phases. The reference air mass 1.5 solar spectral irradiance is shown in yellow (gray solid line), whereas the absorption of $d$-Si is shown by the solid black line.}
\centering
\setlength{\unitlength}{1cm}
{\includegraphics[width=0.99\textwidth]{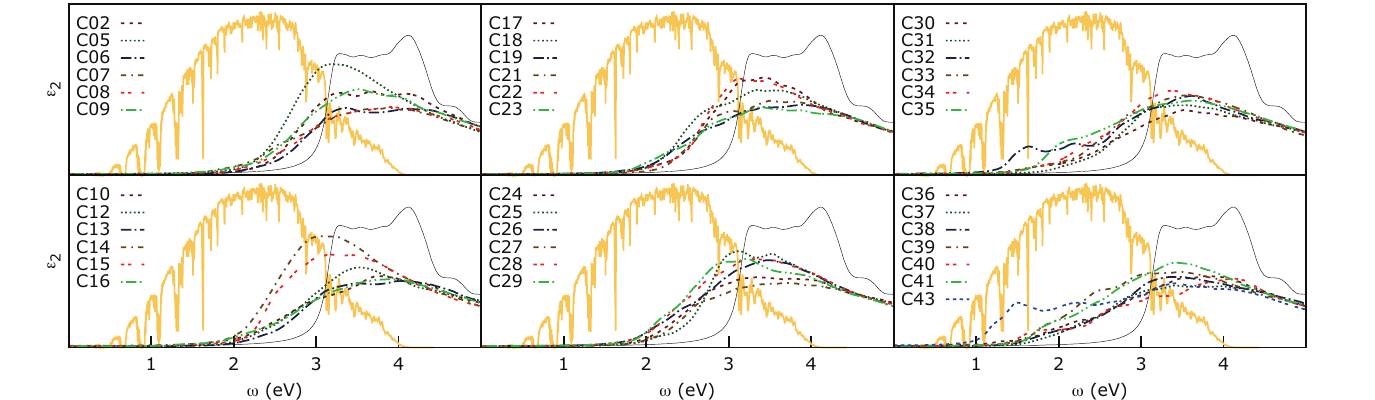}}
\label{fig:spectra}
 \end{figure*}

In Figure~\ref{fig:spectra} the calculated absorption spectra of some of the most promising allotropes are shown. The calculated absorption spectrum of the cubic diamond phase is also shown for comparison, together with the reference air mass (AM) 1.5 solar spectral irradiance~\cite{AM1.5}. We see that, contrary to $d$-Si, most of these new allotropes absorb well in the visible, in a region perfectly compatible with the solar spectrum. This is true even if only a few structures have a direct or quasi-direct band gap. The absorption in this region is still below the ones of the Si20T~\cite{xiang_towards_2013} and P-1~\cite{botti_low-energy_2012} phases proposed before, but still sufficient to allow for thin-film Si photovoltaic cells.

\section{Conclusion}

In conclusion, a systematic search was performed  using a modified version of the minima hopping method, tailored to imitate the synthesis process of silicon clathrates, to identify low density silicon allotropes. The efficiency of our method was confirmed by the discovery of many phases reported in literature and a plethora of hitherto unknown phases. The two main structural motifs leading to low density are the formation of channels or cages. Due to the large sizes of the voids both in the channels and cages they can readily host guest atoms which can be used to tune the material properties. Furthermore, such guest-host materials are likely to be stabilized by pressure as demonstrated by Kurakevych~\textit{et al.}~\cite{kurakevych_na-si_2013}. In view of the potential use in optical applications the electronic properties were investigated by computing the band structure and the absorption spectra. Several low-density allotropes show a significantly improved  overlap with the solar spectrum as compared to the conventional diamond silicon and are thus promising candidates for use in thin-film photovoltaic applications.

This work was done within the NCCR Marvel project. MA gratefully acknowledges support from the Novartis Universit\"{a}t Basel Excellence Scholarship for Life Sciences. We acknowledge the
computational resources provided by the Swiss National Supercomputing
Center (CSCS) in Lugano (project s499) and GENCI (project x2014096017) in France.

%

\end{document}